\documentclass[aps,prx,twocolumn,showpacs,psfig,superscriptaddress,longbibliography]{revtex4-2}

\usepackage{times}
\usepackage{graphicx}
\usepackage{float}
\usepackage{latexsym,amsmath,amssymb,bm,euscript}
\usepackage{color}
\usepackage{subfigure}
\usepackage{epstopdf}
\usepackage[colorlinks=true,linkcolor=blue,citecolor=blue]{hyperref}
\usepackage{soul}
\usepackage[normalem]{ulem}
\usepackage{mathrsfs}
\usepackage{amsmath}
\usepackage{lettrine}
\usepackage{bbding}
\usepackage{xspace}
\usepackage{textcomp}
\usepackage{textcase}
\usepackage{setspace}

\newcommand{\Eq}[1]{Eq.~\eqref{#1}}

\newcommand{\Fig}[1]{Fig.~\ref{#1}}
\newcommand{\Figs}[1]{Figs.~\ref{#1}}

\graphicspath{{./}}
\begin{document}

\title{Spin-Triplet Pairing Induced by Near-Neighbor Attraction in the Cuprate Chain}

\author{Dai-Wei Qu}
\affiliation{CAS Key Laboratory of Theoretical Physics, 
Institute of Theoretical Physics, Chinese Academy of Sciences, 
Beijing 100190, China}

\author{Bin-Bin Chen}
\affiliation{Department of Physics and HKU-UCAS Joint Institute of
Theoretical and Computational Physics, The University of Hong Kong,
Pokfulam Road, Hong Kong, China}

\author{Hong-Chen Jiang}
\affiliation{Stanford Institute for Materials and Energy Sciences, 
SLAC National Accelerator Laboratory and Stanford University, 
Menlo Park, California 94025, USA}

\author{Yao Wang}
\email{yaowang@g.clemson.edu}
\affiliation{Department of Physics and Astronomy, 
Clemson University, Clemson, SC 29631, USA}

\author{Wei Li}
\email{w.li@itp.ac.cn}
\affiliation{CAS Key Laboratory of Theoretical Physics, 
Institute of Theoretical Physics, Chinese Academy of 
Sciences, Beijing 100190, China}
\affiliation{CAS Center for Excellence in Topological 
Quantum Computation, University of Chinese Academy of 
Sciences, Beijing 100049, China}

\begin{abstract} 
\end{abstract}
\date{\today}
\maketitle

\noindent{\bf{Abstract}}\\
{In quantum materials, the electronic interaction and the electron-phonon 
coupling are, in general, two essential ingredients, the combined impact of 
which may drive exotic phases. Recently, an anomalously strong 
electron-electron attraction, mediated by phonons, has been unveiled 
in one-dimensional copper-oxide chain Ba$_{2-x}$Sr$_x$CuO$_{3+\delta}$.
Yet, it is unclear how this 
strong near-neighbor attraction $V$ influences the superconductivity pairing 
in the compound. Here we perform accurate many-body calculations to study 
the extended Hubbard model with on-site Coulomb repulsion $U>0$ and 
attraction $V<0$ that well describes the cuprate chain and likely other similar 
transition-metal materials with both strong correlations and lattice effects. We find a rich quantum phase diagram containing 
an intriguing Tomonaga-Luttinger liquid phase --- besides the spin density 
wave and various phase separation phases --- that can host dominant spin-triplet 
pairing correlations and divergent superconductive susceptibility. Upon doping, 
the spin-triplet superconducting regime can be further broadened in the 
parameter space and extends to larger $U$, offering a feasible mechanism 
to realize $p$-wave superconductivity in realistic cuprate chains.
\\
}

\noindent{\bf{Introduction}}\\
Strongly correlated materials, where the electronic structure cannot be
approximated by the reductive band theory, have become a research 
frontier. In particular, two types of unconventional superconductivity 
have attracted considerable attention. One of them is the high-$T_c$ 
superconductivity discovered in cuprates~\cite{Cuprate1986}. Although 
this class of materials has been investigated for nearly 40 years, 
the pairing mechanism remains an enigma~\cite{Keimer2015Nature,
Zhou2021Review}. The other type of unconventional superconductivity 
is the topological triplet-pairing superconductivity~\cite{Sato2017Review,
QiXiaoLiang2011RMP,Hasan2010RMP}, where electron fractionalizes 
into Majorana excitations~\cite{Kitaev2001,read2000paired} and is the 
foundation for topological quantum computing~\cite{Stern2013Science,
Nayak2008RMP}. Therefore, pursuing such exotic superconductivity 
in realistic compounds constitutes a stimulating research topic.

The single-band Hubbard model, as the prototypical model carrying the 
strong correlation effects, has been widely employed in the studies of 
many-body electron systems~\cite{Wietek2021METTS,Qin2020Absence,YFJiang2020PRR,
HCJiang2019Science,Zheng2017Sci,LeBlanc2015} as variants of this model are 
relevant to the two-dimensional (2D) cuprate superconductors. Besides, 
quasi-1D cuprate chains also constitute important class of strongly 
correlated materials that host intriguing correlated electron states and 
effects, e.g., the Tomonaga-Luttinger liquid (TLL) with spin-charge 
separation~\cite{Kim1996PRL,Fujisawa1999PRB,Kim2006NatPhys}. 
On the other hand, most theoretical studies of the ground-state
and dynamical properties~\cite{Tomita2001,Benthien2007,Hofmann2012,
Al-Hassanieh2013} also lie in 1D as rigorous many-body simulations are 
more accessible using analytics~\cite{Essler2010HubbardChain}, exact 
diagonalization, density matrix renormalization group (DMRG)~\cite{White1992} 
and quantum Monte Carlo~\cite{Hirsch1983,Hirsch1984a,Hirsch1984b}. Since 
both the on-site interaction $U$ and near-neighbor (NN) interaction $V$ 
correspond to the electronic repulsion at different distances, previous 
numerical studies focused on the cases with repulsive $U,V>0$ as supposed 
relevant to real materials~\cite{Hirsch1984PRL,Jeckelmann2002PRL,
Sandvik2004PRL,Gu2004PRL,Ejima2007PRL}.

Most recently, a paradigm shift occurs as an \emph{in situ} ARPES 
experiment on the 1D cuprate chain Ba$_{2-x}$Sr$_x$CuO$_{3+\delta}$ 
(BSCO) has revealed an anomalously strong attraction $V<0$ between NN 
electrons\,\cite{Chen2021CuprateChain}. In contrast to the intrinsic 
electron-electron Coulomb repulsion, this attractive interaction is likely 
to be mediated by the strong electron-phonon coupling in transition 
metal oxides~\cite{Wang2021AttractivePRL}. Such an effective attraction 
largely missed previously may serve as a key ingredient in both 
understanding the high-$T_\textrm{c}$ superconductivity and enabling 
exotic quantum phases in correlated materials~\cite{Lin1986PRB,
mila1993phase,penc1994phase,lin1995phase,Lin1997PhyC,
Nakamura_PRB2000,xiang2019doping}. Therefore, an interesting 
question naturally arises: Does such an effective attraction $V$ help 
establish superconductivity pairing between the strongly correlated electrons? 

\begin{figure}[!tbp]
\includegraphics[width=1\linewidth]{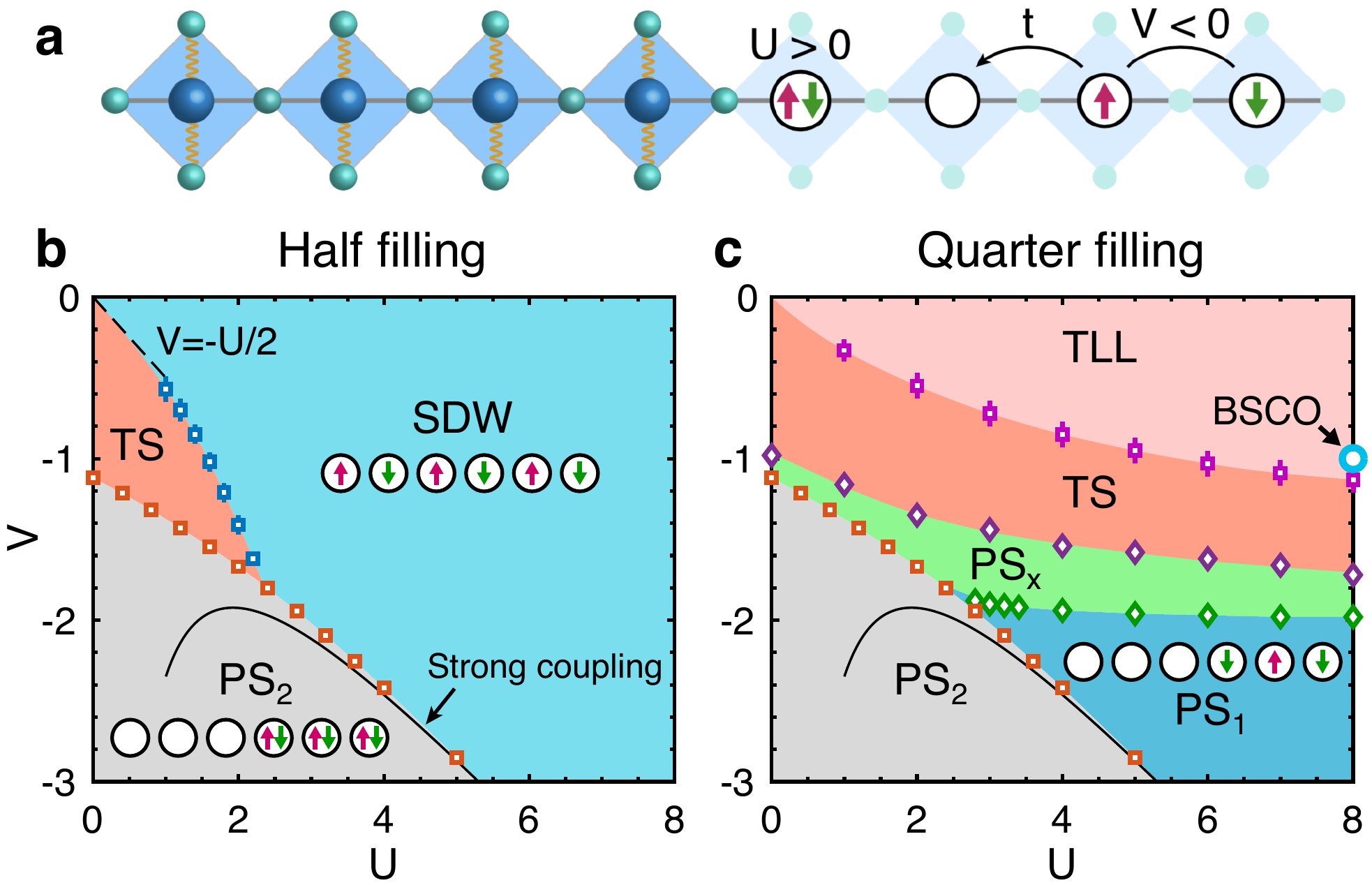}\vspace{-6mm}
\renewcommand{\figurename}{\textbf{Fig.}}
\caption{\textbf{Extended Hubbard model and phase diagrams.}
\textbf{a} illustrates the BSCO compound and corresponding extended 
$t$-$U$-$V$ Hubbard model with NN hopping $t$, on-site repulsive $U$, 
and NN attractive $V$ terms. \textbf{b} and \textbf{c} show the quantum 
phase diagrams of the EHM at half and quarter fillings, 
respectively. The solid black line represents the 
asymptotic phase boundary $V=-\frac{U}{2}-\frac{8\ln2}{3U}$ 
in the strong coupling limit, and the dashed line for 
$V=-U/2$ in the weak coupling limit~\cite{Lin1986PRB,Nakamura_PRB2000}. 
The blue circle in \textbf{c} represents the parameters $U=8$ 
and $V=-1$ of the doped 1D cuprate chain BSCO~\cite{Chen2021CuprateChain}.
}
\label{Fig:PhDiag}
\end{figure}

To address this question, and also motivated by the recent experimental 
realization of such attractive-$V$ extended Hubbard model (EHM, 
see Fig.~\ref{Fig:PhDiag}\textbf{a}), we employ large-scale DMRG 
simulations and systematically explore its phase diagram. We 
especially focus on the possible realization of spin-triplet 
superconductivity while identifying all phases. At both half and 
quarter fillings, we have numerically determined the ground-state 
phase diagrams of the EHM, from which we identify a robust 
gapless TLL phase with a prominent spin-triplet superconducting 
pairing (TS) with algebraic singularity. In two dimensions, the 
triplet superconducting (SC) state is topologically non-trivial 
where the fractional excitation can emerge on the boundary
\cite{read2000paired,sarma2006proposal,bolech2007observing,
tewari2007index,sau2010robustness}. However, quantum fluctuations 
are usually too strong in 1D such that interacting electrons in 
a Hubbard-type chain usually behave as a TLL, contradicting 
the mean-field and small-cluster predictions. Therefore in 
this paper, we refer this emergent TLL phase with divergent 
superconducting susceptibility to as \textit{a gapless TS phase}.

Our main findings are summarized in Fig.~\ref{Fig:PhDiag}. At half filling 
(see Fig.~\ref{Fig:PhDiag}\textbf{b}), the TS phase survives only up to a 
finite $U_\textrm{c}/t\simeq 2.3$ and is absent when $U>U_\textrm{c}$. 
At quarter filling (see Fig.~\ref{Fig:PhDiag}\textbf{c}), this TS 
phase extends to larger $U$s comparable to those in cuprates
\cite{Chen2021CuprateChain}. Between this TS phase and 
the regular PS phases with singly (PS$_1$) and doubly (PS$_2$) 
occupied clusters, we further identify an exotic PS$_x$ phase where 
the clustered electrons form the TLL and even TS states. With the 
model parameters determined from fitting dynamical data of BSCO, 
our study reveals a close proximity of this doped cuprate chain to the 
$p$-wave superconductivity, and provide theoretical guide for realizing 
such gapless TS phase in 1D cuprate chains. 
\\

\noindent{\bf{Results}}\\
\textbf{EHM with NN attraction.}
The BSCO chain can be described by the EHM with on-site $U>0$ 
and NN attraction $V<0$, whose Hamiltonian reads 
\begin{equation}
 H =  -t\sum_{i=1, \sigma}^{L-1} ( c_{i\sigma}^\dag c_{i+1 \sigma}^{\,} + \text{H.c.} )  + U\sum_{i=1}^L n_{i\uparrow} n_{i\downarrow}
       + V\sum_{i=1}^{L-1}  n_i n_{i+1} ,
\label{Eq:Ham}
\end{equation}
where $c_{i\sigma}^\dag$ ($c_{i \sigma}$) is the electron creation (annihilation)
operator, $\sigma=\uparrow,\downarrow$ labels the electron spin, and $n_i = n_{i\uparrow} + n_{i\downarrow}$ is the particle number operator at site $i$. 
Throughout the study, we set hopping amplitude $t=1$ as the energy unit, and 
focus on the ground state phase diagrams at both half and quarter fillings. In 
this work, we employ DMRG method with non-Abelian symmetry implemented
\cite{Weichselbaum2012,Weichselbaum2020} (see Methods and Supplementary Note~1).

To characterize various quantum phases, we compute the spin, charge, 
and pairing correlation functions. The spin-spin correlation is defined as
$F(r) = \langle \mathbf{S}_{i} \cdot \mathbf{S}_{j} \rangle,$ 
with $\mathbf{S}_{i(j)}$ the spin operator at site $i(j)$ and $r \equiv j-i$. 
The charge density correlation is defined as $D(r) = \langle n_i n_j \rangle 
- \langle n_i \rangle \langle n_j \rangle,$ where $n_{i(j)}$ is the particle 
number operator at site $i(j)$. To characterize the superconducting pairing 
correlation, we consider both the spin-singlet ($s$-wave) pairing 
$\Phi_\textrm{S} (r) = \langle \Delta_\textrm{S}^\dag (i) \Delta_\textrm{S} (j) 
\rangle$ with $\Delta^\dag_\textrm{S}(i) = \frac{1}{\sqrt{2}} (c^\dag_{i,\uparrow}
c^\dag_{i+1,\downarrow} - c^\dag_{i,\downarrow} c^\dag_{i+1,\uparrow})$,
and the triplet ($p$-wave) pairing $\Phi_{\textrm{T},s} (r) = 
\langle \Delta_{\textrm{T},s}^\dag (i) \Delta_{\textrm{T},s} (j) \rangle$
with three components $\Delta^\dag_{\textrm{T},1}(i) 
= c^\dag_{i,\uparrow}c^\dag_{i+1,\uparrow}$, 
$\Delta^\dag_{\textrm{T},0}(i) = \frac{1}{\sqrt{2}} 
( c^\dag_{i,\uparrow} c^\dag_{i+1,\downarrow} + c^\dag_{i,\downarrow} 
c^\dag_{i+1,\uparrow} )$, and $\Delta^\dag_{\textrm{T},-1}(i) 
= c^\dag_{i,\downarrow}c^\dag_{i+1,\downarrow}$ 
for $s=1,0,-1$, respectively. Note that the EHM in Eq.~(\ref{Eq:Ham}) 
is SU(2) invariant so the above three components are degenerate in 
the spin-triplet channel, and we thus take the {averaged $\Phi_\textrm{T}(r) 
= \frac{1}{3} \sum_s \Phi_{\textrm{T},s}(r)$} from our SU(2) DMRG 
calculations and compare it with $\Phi_\textrm{S}$.
\\

\noindent
\textbf{Analytical results from the TLL theory.} The TLL theory puts rigorous 
constraints~\cite{Haldane1981Luttinger,Giamarchi1D,FradkinFieldTheory,
Voit_RepProgPhys1995} on our numerical results, which we always compare 
with and make use of in the analysis of our numerical data. In TLL, two-point 
correlation functions including the spin, charge and pairing correlations all 
decay in power law $\sim r^{-\alpha}$, with exponents $\alpha$ determined 
by two basic Luttinger parameters $K_\sigma$ and $K_\rho$, respectively 
related to the spin and charge degrees of freedom (see more details in the 
Supplementary Note~2). To accurately evaluate these intrinsic parameters, 
one can calculate the momentum-dependent spin structure factor 
$S_\textrm{m}(k)$ and charge structure factor $S_\textrm{c}(k)$, and then 
extract $K_\sigma$ and $K_\rho$. 

For the current EHM in Eq.~(\ref{Eq:Ham}) with SU(2) spin symmetry, 
$K_\sigma=1$ for the spin density wave (SDW), TLL, and TS phases with 
gapless spin excitations, while $K_\sigma=0$ in the spin gapped phase 
PS$_2$. Therefore, $K_\rho$ uniquely determines the power-law exponents 
$\alpha$ of various correlations: for charge and spin correlations there exist 
a uniform mode with exponent $\alpha_0=2$ and a 2$k_F$ mode with 
$\alpha_{2k_F} =1+K_\rho$; for the pairing correlations $\Phi_\textrm{S}$ 
and $\Phi_\textrm{T}$, {they both have uniform modes with the same exponent} 
$\alpha_\textrm{SC} = 1+1/K_\rho$, which dominates over the spin and charge 
correlations when $K_\rho>1$. Consequently,  the low-$T$ behaviors of the 
staggered magnetic, charge, and pairing susceptibilities are also controlled 
by $K_\rho$, i.e., $\chi_\textrm{SDW} \sim {T^{K_\rho-1}}$, 
$\chi_\textrm{CDW} \sim {T^{K_\rho-1}}$, 
and $\chi_\textrm{SC} \sim {T^{1/K_\rho-1}}$.
For $K_\rho > 1$ or $<1$, these susceptibilities 
exhibit apparently distinct behaviors as $T\to0$.
Thus, the Luttinger parameter constitutes an essential 
quantity characterizing the underlying phases of a 1D system.
In practice, we extract the Luttinger parameter $K_\rho$ via 
a second-order polynomial fitting of $S_\textrm{c}(k)$ 
in the small $k$ regime~\cite{Giamarchi1D,Sandvik2004PRL,
Ejima2007PRL,PRB2011_1DtJ} (see Supplementary Note~3 
for details). To minimize the boundary effect, we evaluate the 
correlation functions using sites away from both ends.
\\

\begin{figure}[!tbp]
\includegraphics[width=1\linewidth]{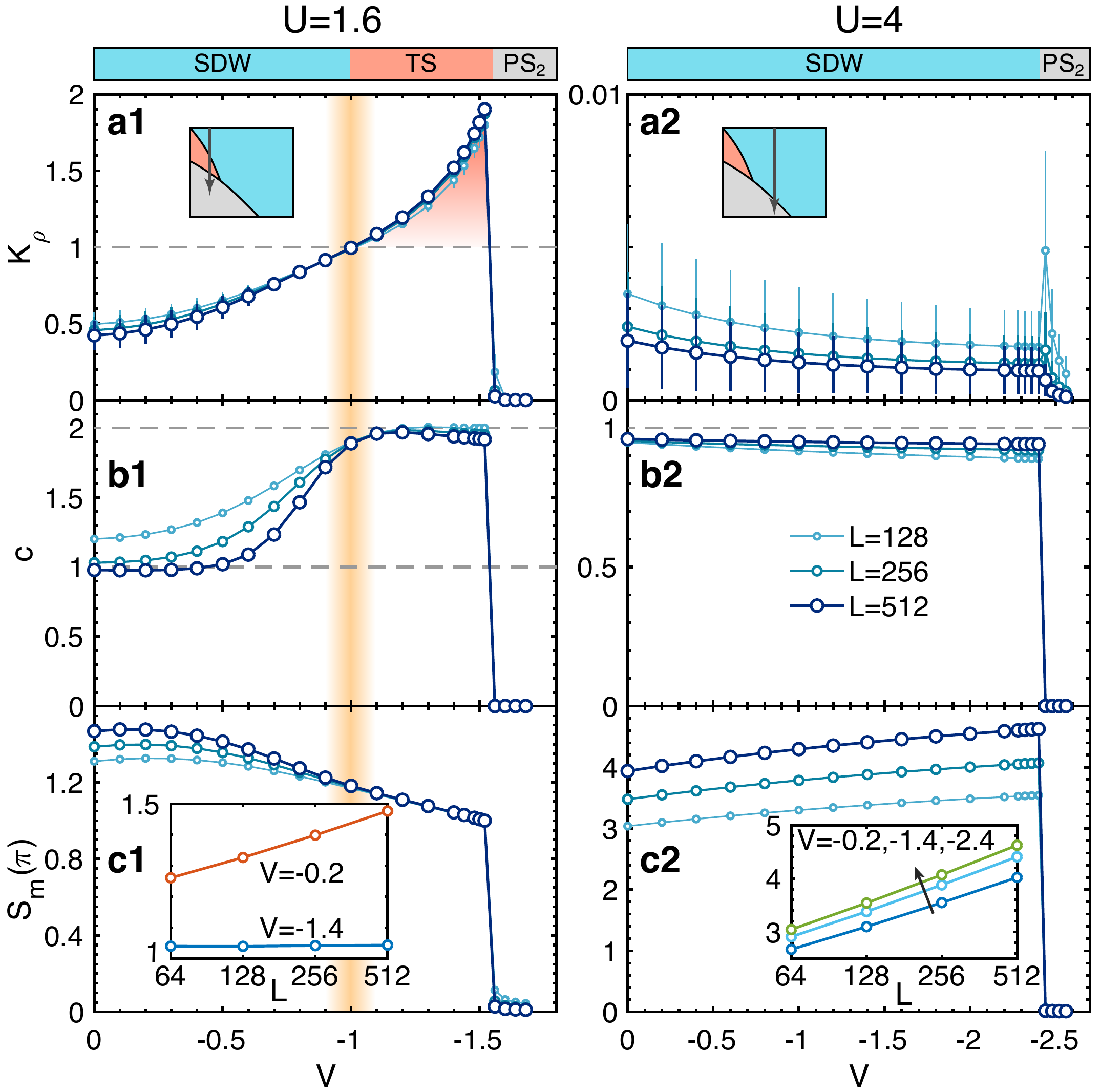}\vspace{-6mm}
\renewcommand{\figurename}{\textbf{Fig.}}
\caption{\textbf{Results of EHM at half filling.} \textbf{a1}-\textbf{c1} 
are with on-site repulsion of $U=1.6$ and \textbf{a2}-\textbf{c2} of $U=4$, 
with the corresponding mini phase diagrams also depicted. \textbf{a1},
\textbf{a2} show the results of the Luttinger parameter $K_\rho$, 
\textbf{b1}, \textbf{b2} are the results of central charge $c$ determined 
from the entanglement entropy scalings,  and \textbf{c1}, \textbf{c2} are 
the spin structure factors at $\pi$. Here the $K_\rho$ results in 
\textbf{a1}, \textbf{a2} are obtained via the second-order polynomial 
fitting of $S_\textrm{c}(k)$ in the small-$k$ regime (Supplementary Note~3). 
The insets in \textbf{c1}, \textbf{c2} show $S_\textrm{m}(\pi)$ vs. $L$ 
at various $V$s. Panels \textbf{a1}-\textbf{c1} reveal the system 
undergoes a transition from SDW to TS phase at $V_\textrm{c}\simeq -1$ 
and then enters the PS$_2$ regime for $V < V_\textrm{s} \simeq -1.55$. 
Panels \textbf{a2}-\textbf{c2} show no intermediate phase but a direct 
first-order transition from SDW to PS$_2$ at $V_\textrm{s}\simeq-2.42$.
}
\label{Fig:TS}
\end{figure}

\begin{figure*}[!tbp]
\includegraphics[width=1\linewidth]{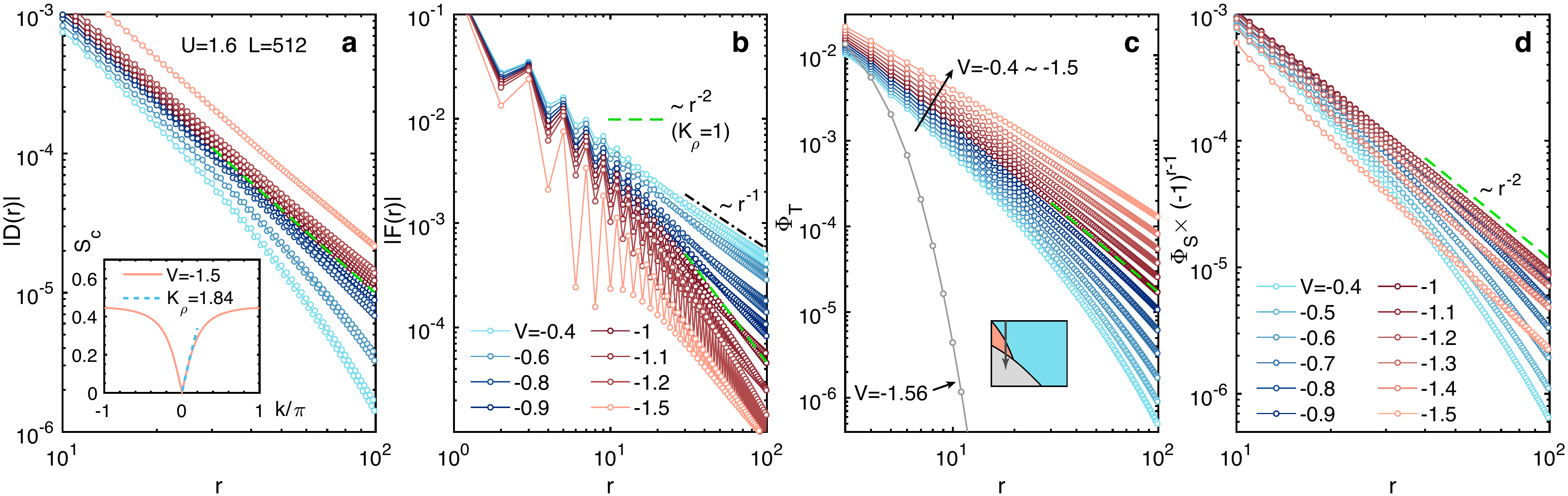}\vspace{-6mm}
\renewcommand{\figurename}{\textbf{Fig.}}
\caption{
\textbf{Correlation functions of the half-filled EHM at $U=1.6$.}
Here we show \textbf{a} density-density, \textbf{b} spin-spin, 
\textbf{c} the spin-triplet pairing, and \textbf{d} singlet 
pairing correlations, with various attractive interactions $-1.5\leq V\leq 
-0.4$. The color codes in the four panels are the same (following the
representative color of each phase in Fig.~\ref{Fig:PhDiag}\textbf{b}, 
as indicated by the legends in \textbf{b}, \textbf{d}).
The dashed green line represents $\sim r^{-2}$ scaling 
that corresponds to $K_\rho=1$, separating the exponential 
and power-law behaviors of various correlations in panels 
\textbf{a}, \textbf{c}, \textbf{d}. The dotted dash line $\sim r^{-1}$ with exponent
$\alpha=1$ in \textbf{b} represents long-distance scaling of the spin
correlations in the SDW phase. The inset in \textbf{a} shows the
charge structure factor at $V=-1.5$, whose second-order
polynomial fitting (dashed cyan line) results
in a Luttinger parameter of $K_\rho=1.84$. 
}
\label{Fig:Pairing}
\end{figure*}

\noindent{\textbf{Quantum phase diagram at half filling.}}
We summarize our main findings at half filling in the phase 
diagram of Fig.~\ref{Fig:PhDiag}\textbf{b}, where the SDW, 
phase separation PS$_2$ with doubly occupied sites clustered, 
and most remarkably, a TLL phase with prominent superconductive
pairing is uncovered. To show the distinction of these phases, 
we present simulations along two typical paths in Fig.~\ref{Fig:TS}, 
namely, the $U=1.6$ and $U=4$ vertical cuts in the phase diagram.

The Luttinger parameter $K_\rho$ clearly separates the $U=1.6$ 
systems into three regimes. As the interaction strength increases 
to $|V|>|V_\textrm{c}|\simeq1$ (but smaller than the phase separation 
transition strength $|V_\textrm{s}|$, which will be discussed later), 
in Fig.~\ref{Fig:TS}\textbf{a1} there exists an intermediate regime 
with $K_\rho>1$. We also compute the central charge $c$ by fitting 
the entanglement entropy (see more details in Supplementary Note~4), 
and from Fig.~\ref{Fig:TS}\textbf{b1} $c$ is found to change from 
$c\simeq 1$ to about 2 for $|V_\textrm{c}| < |V| < |V_\textrm{s}|$, 
confirming that the intermediate phase has both gapless spin and 
charge modes. On the other hand, also as shown in Fig.~\ref{Fig:TS}, 
for the $U=4$ case $K_\rho$ remains small for all values of $V$ 
and does not exceed 1 (see Fig.~\ref{Fig:TS}\textbf{a2}) and the 
central charge remains $c=1$ (Fig.~\ref{Fig:TS}\textbf{b2}), 
showing the absence of such intermediate phase.

With further increase of the attractive interaction for 
either $U=1.6$ or 4, the system eventually exhibits 
phase separation for $|V| > |V_\textrm{s}|$. The critical strength 
$V_\textrm{s}$ dependent on $U$ is shown in Fig.~\ref{Fig:PhDiag} 
(see the detailed estimation of $V_\textrm{s}$ in 
Supplementary Note~1). Specifically for the two 
selected cuts, we found  $V_\textrm{s}\simeq -1.55$ for $U=1.6$ 
(see Fig.~\ref{Fig:TS}\textbf{a1}-\textbf{c1}) and $V_\textrm{s}\simeq -2.42$ for 
$U=4$ (see Fig.~\ref{Fig:TS}\textbf{a2}-\textbf{c2}). In such a PS state,
the clustered part consists of doubly-occupied sites and 
no singularity can be observed in various correlations. 
Therefore, we denote it as PS$_2$ to distinguish from 
other PS phases discussed later.

Among these three phases in the $U=1.6$ case (and 
for other interactions $U < U_\textrm{c} \simeq 2.3$, c.f.,
Fig.~\ref{Fig:PhDiag}\textbf{b}), we are particularly interested 
in the intermediate one due to the signature of triplet 
pairing. As evidenced by the charge correlation results in
Fig.~\ref{Fig:Pairing}\textbf{a}, the charge gap is closed by 
the attractive $V$ term, and the Luttinger parameter $K_\rho$ 
can be fitted to be greater than 1 (see the inset of
Fig.~\ref{Fig:Pairing}\textbf{a}, and more details in 
Supplementary Note~3). 
According to the TLL theory, the
superconductive paring decays $r^{-\alpha_\textrm{SC}}$
with the exponent $\alpha_\textrm{SC}=1+1/K_\rho$---
smaller than the algebraic exponent ($1+K_\rho$) of 
both the charge and spin correlations when $K_\rho>1$---
and thus constitutes the dominant correlation in the charge-2e channel, 
with an algebraically diverging pairing susceptibility
$\chi_\textrm{SC}(T)$ for low temperature $T$.

In the weak attraction regime $|V|<|V_\textrm{c}|$, $K_\rho$ 
vanishes in the thermodynamic limit \footnote{Note that 
due to the strong finite-size effects and a small charge 
gap for $U=1.6$, in the weak coupling $|V|<|V_\textrm{c}|$ regime, 
$K_\rho$ remains finite and it converges to zero only in 
the thermodynamic limit.} and $K_\sigma=1$ due to the spin 
SU(2) symmetry. In Fig.~\ref{Fig:Pairing}\textbf{b}, an quasi-long 
range spin order with an algebraic exponent of
$\alpha_\textrm{SDW}=1$ appears, which has logarithmically 
diverging spin structure factor of $S_\textrm{m}(k=\pi)$ 
(see Fig.~\ref{Fig:TS}\textbf{c1},\textbf{c2} and the insets). This is 
well consistent with the SDW scenario with a finite charge gap and
quasi-long range spin order (see Supplementary Note~5). On the 
other hand, for the intermediate phase in Fig.~\ref{Fig:TS}\textbf{c1}
$S_\textrm{m}(\pi)$ ceases to increase vs. $L$, as the 2$k_F$ 
mode spin correlation decays faster than $\sim r^{-2}$ shown 
in Fig.~\ref{Fig:Pairing}\textbf{b}, which reveals a non-diverging 
magnetic susceptibility and thus rather distinct magnetic 
properties from that of the SDW phase. 
\\

\begin{figure}[!tbp]
\includegraphics[width=1\linewidth]{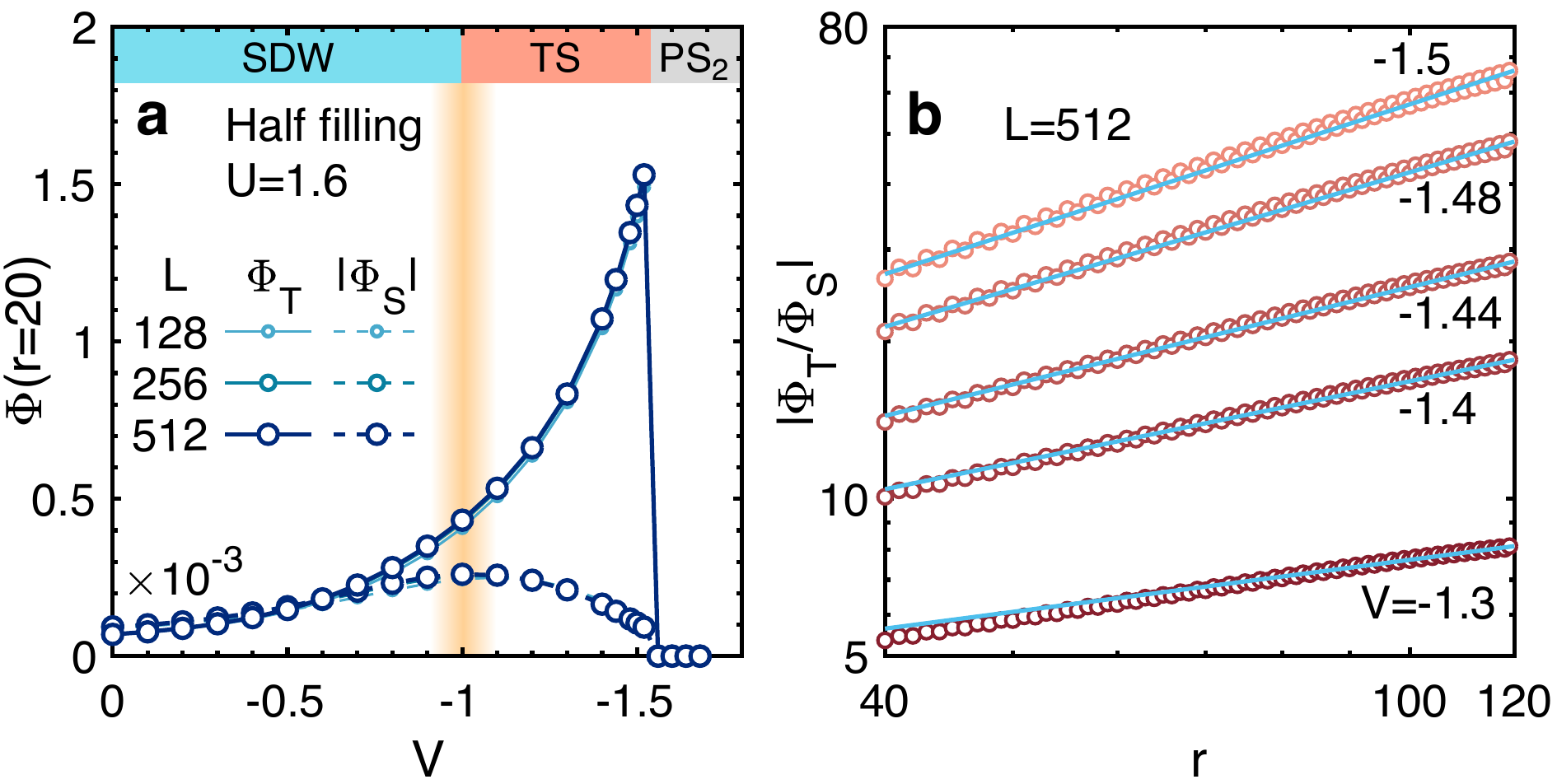}\vspace{-6mm}
\renewcommand{\figurename}{\textbf{Fig.}}
\caption{\textbf{Dominance of the triplet over singlet pairing in 
the half-filled TS phase.} Here we show the data of $U=1.6$.
\textbf{a} Singlet and triplet pairing correlations at a fixed 
(long) distance $r=20$. The yellow strip represents the SDW-TS 
transition point $V_\textrm{c}\simeq -1$. \textbf{b} The calculated
ratios $|\Phi_\textrm{T}/\Phi_\textrm{S}|$, plotted in a log-log scale, 
exhibit excellent agreement with the scaling $r^{K_\rho-1}$ 
(blue solid lines).
}
\label{Fig:Triplet}
\end{figure}

\noindent{\textbf{Gapless triplet superconducting phase.}}
As shown in Fig.~\ref{Fig:Pairing}\textbf{c},\textbf{d}, it can be 
observed that both the singlet- ($\Phi_\textrm{S}$) and 
triplet-pairing ($\Phi_\textrm{T}$) exhibit power-law decay 
behaviors, and the latter with $p$-wave pairing symmetry clearly 
dominates over the former with the $s$-wave pairing symmetry. 
This is clearly demonstrated in Fig.~\ref{Fig:Triplet}\textbf{a}, 
where the strengths of the two correlations $\Phi_\textrm{T}(r)$ 
and $\Phi_\textrm{S}(r)$ are compared at a fixed distance $r=20$.
Though two pairing correlations are comparable in the SDW regime,
$\Phi_\textrm{T}(r)$ clearly surpasses $\Phi_\textrm{S}(r)$ 
once entering the intermediate-$V$ phase: the latter turns 
to decreasing, while $\Phi_\textrm{T}(r)$ keeps increasing 
and becomes over one order of magnitude greater than
$\Phi_\textrm{S}(r)$.

Such a dominance of the triplet pairing in the TS phase holds 
for different distances $r$ other than the fixed distance $r=20$ 
in Fig.~\ref{Fig:Triplet}\textbf{a}. This dominance is reflected in 
the spatial distribution of both pairing correlations in 
Fig.~\ref{Fig:Pairing}\textbf{c},\textbf{d}. There we find 
$\Phi_\textrm{T}$ firstly decay exponentially in the SDW 
phase (the blue dots), then exhibits power-law behaviors for 
$|V_\textrm{c}| < |V| < |V_\textrm{s}|$ (the red dots), and 
decays again exponentially for $|V|>|V_\textrm{s}|$ (the grey 
dots). We notice there is virtually no uniform but only $2k_F$ 
mode in $\Phi_\textrm{S}$, as reflected in the smooth curves 
$\Phi_\textrm{S}(r) \times (-1)^{r-1}$ in Fig.~\ref{Fig:Pairing}\textbf{d}. 
For the gapless TS phase where we are most interested in, the 
dominance of $\Phi_\textrm{T}$ is reflected by the comparison of 
Figs.~\ref{Fig:Pairing}\textbf{c} and \textbf{d}: $\Phi_\textrm{T}(r)$ 
decays slower than $r^{-2}$, while $\Phi_\textrm{S}(r)$ decays faster 
than $r^{-2}$ (Fig.~\ref{Fig:Pairing}\textbf{d}). More quantitatively,
the ratio between these two pairing correlations $|\Phi_\textrm{T}(r)/
\Phi_\textrm{S}(r)|$ scales in power law $r^{K_\rho-1}$, since the 
leading scaling in $\Phi_\textrm{T}$ and $\Phi_\textrm{S}$ is 
$1/r^{1+1/K_\rho}$ and $1/r^{K_\rho+1/K_\rho}$, respectively 
(see Supplementary Note~2). We present such a power-law 
scaling extracted from our DMRG simulations in Fig~\ref{Fig:Triplet}\textbf{b}. 
Therefore, in the intermediate regime the pairing correlation $\Phi_\textrm{T}$ 
dominates over $\Phi_\textrm{S}$ not only in magnitude but actually 
in long-distance scaling, making it a rather unique gapless TS phase.

\begin{figure*}[!tbp]
\includegraphics[width=1\linewidth]{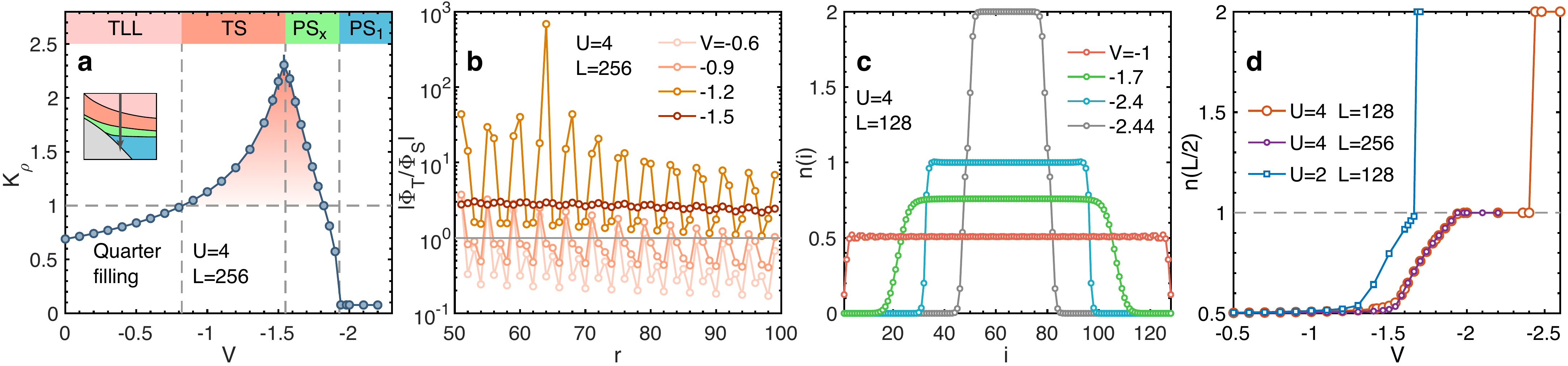}\vspace{-6mm}
\renewcommand{\figurename}{\textbf{Fig.}}
\caption{\textbf{Results of EHM at quarter filling.} We show in 
\textbf{a} the mini phase diagram and Luttinger parameter $K_\rho$, 
and \textbf{b} accordingly the ratio between triplet and singlet 
pairings $|\Phi_\textrm{T}/\Phi_\textrm{S}|$ that becomes greater 
than 1 in the emergent TS phase. The $K_\rho$ results in the 
PS$_x$ and PS$_1$ are computed within the clustered part 
(see Supplementary Note~3). \textbf{c} The charge distribution 
$n(i)$ in TS ($V=-1$), PS$_x$ ($V=-1.7$), PS$_1$ ($V=-2.4$), 
and PS$_2$ ($V=-2.44$) phases, respectively. \textbf{d} The 
charge densities $n(L/2)$ measured at the center of the system 
are plotted versus $V$. 
}
\label{Fig:PS}
\end{figure*}

When compared to the phase diagram obtained in 
Ref.~\onlinecite{Lin1986PRB}, our DMRG results in
Fig.~\ref{Fig:PhDiag}\textbf{b} show some agreement on the 
existence of three phases, yet there are still noticeable differences. 
Particularly, our DMRG calculations identify the upper 
boundary of the TS phase in agreement with $V=-U/2$ 
obtained from the perturbation theory in the small $U$ regime 
while it deviates from this line in the strong coupling regime. 
Consequently, in contrary to Ref.~\onlinecite{Lin1986PRB} where
the TS phase was shown extending to infinite $U$, our results 
in Fig.~\ref{Fig:PhDiag}\textbf{b} suggest it can only survive 
up to $U_\textrm{c}\simeq 2.3$, located in a much narrower 
regime. On the other hand, when compared to more recent studies~\cite{Iemini2015,mendozaarenas2021dynamical} 
where the phase diagrams are only schematic, here we 
pinpoint the numerically accurate phase boundaries with
large-scale DMRG calculations and reveal the predominant 
triplet quasi-long range TS pairing relevant to the realistic 
cuprate chain BSCO, decades after such a TS instability was proposed~\cite{Lin1986PRB,Lin1997PhyC}.
\\

\noindent{\textbf{Finite doping.}}
Besides half filling, we have also explored the phases in the 
doped EHM systems. We first focus on the quarter filling, where 
the triplet pairing instability is approximately maximized, as will 
be discussed later. The extracted phase diagram is presented 
in Fig.~\ref{Fig:PhDiag}\textbf{c}. Here, we select a cut along 
$U=4$ and explain the properties of each phase in Fig.~\ref{Fig:PS}. 
Similar to half filling, the Luttinger parameter $K_\rho > 1$ 
characterizes the intrinsic nature of the correlations and 
separates the $U=4$ systems into four phases (see 
Fig.~\ref{Fig:PS}\textbf{a}). Particularly, for $V<V_\textrm{c}\simeq-0.8$, 
we identified a TS regime following the same principle as
half filling, manifested as enhanced triplet and singlet 
pairing correlations. Between these two correlations, we 
evaluated their ratio $|\Phi_\textrm{T}/\Phi_\textrm{S}|$ 
and found its envelop increasing monotonically 
as $|V|$ enhances and exceeding 1 for $V<V_\textrm{c}$ 
(see Fig.~\ref{Fig:PS}\textbf{b}), despite some oscillations 
with distance $r$. Note the two pairing correlations now show
the same scaling at long distance. Importantly, the TS phase at
quarter filling is significantly wider than that at half filling,
particularly in the large $U$ regime.

Besides the TS regime, there are three different inhomogeneous 
PS phases, i.e., PS$_1$, PS$_x$, and PS$_2$ in
Fig.~\ref{Fig:PhDiag}\textbf{c}, in the doped system. 
The real-space charge distributions $n(i)$ are shown 
in Fig.~\ref{Fig:PS}\textbf{c}, from which we see that in 
the PS phases the electrons cluster with filling $n=1, 2$ 
or $x \in (1/2,1]$. To track the evolution among 
these PS phases when $V$ changes, we pick the center of 
the system as a representative, which always lies in the 
filled domain in a PS state due to the open boundary, 
and extract $n(i=L/2)$ for different $U$ and $V$ strengths 
in Fig.~\ref{Fig:PS}\textbf{d}. This filling density starts with
$n(L/2)=0.5$ (i.e., the TLL and TS phases) and deviates from 
the uniform quarter filling when $|V|$ is stronger than 
certain transition value. As $n(L/2)=x$ is not a fixed 
integer value but varies between 0.5 and 1, we denote this 
regime as PS$_x$. For small $U$, like $U=2$, the system
jumps from PS$_x$ to PS$_2$ at a second transition point. 
In contrast, this transition is preceded by a third PS phase 
for large $U>U_\textrm{c}\simeq2.3$ (the same as that of half filling).
Taking $U=4$ as an example, PS$_x$ firstly transits into an 
$n(L/2)=1$ phase (denoted as PS$_1$), and then jumps into PS$_2$ 
as $|V|$ further increases. For the doped cases with filling 
factors other than 1/4, the quantum phase diagram is
qualitatively similar to that of Fig.~\ref{Fig:PhDiag}\textbf{c}. 
The phase boundaries of PS$_1$ and PS$_2$ actually remain 
intact for other doping since they reflect the local energy
relation between singly and doubly occupied states. The 
quantum many-body states in the clustered part of the three 
PS phases --- PS$_1$, PS$_2$, and PS$_x$ --- only depends 
on the interaction parameters $U$ and $V$.

The existence of the PS$_x$ phase was missed in early studies 
on the same model~\cite{Lin1997PhyC,Lin1986PRB}, and the 
distinct feature of PS$_x$ is the clustered electrons that 
constitute a TLL liquid with fractional filling. With $x$ continuous tuned by $V$, the clustered 
part of PS$_x$ can also become close to half filling in terms of density, i.e., 
$x=1$. Nevertheless, it is distinct from that in the PS$_1$ 
phase, as the clustered electrons in the latter form a charge 
gapped SDW instead of a gapless TLL. Even more interestingly, 
we can also identify a $K_\rho>1$ regime and significant TS pairing correlations 
in the clustered part of PS$_x$, showing the existence of gapless
TS cluster in (at least part of) the PS$_x$ phase (see more 
details in Supplementary Note~6). 
\\

\begin{figure}[!tbp]
\includegraphics[width=1\linewidth]{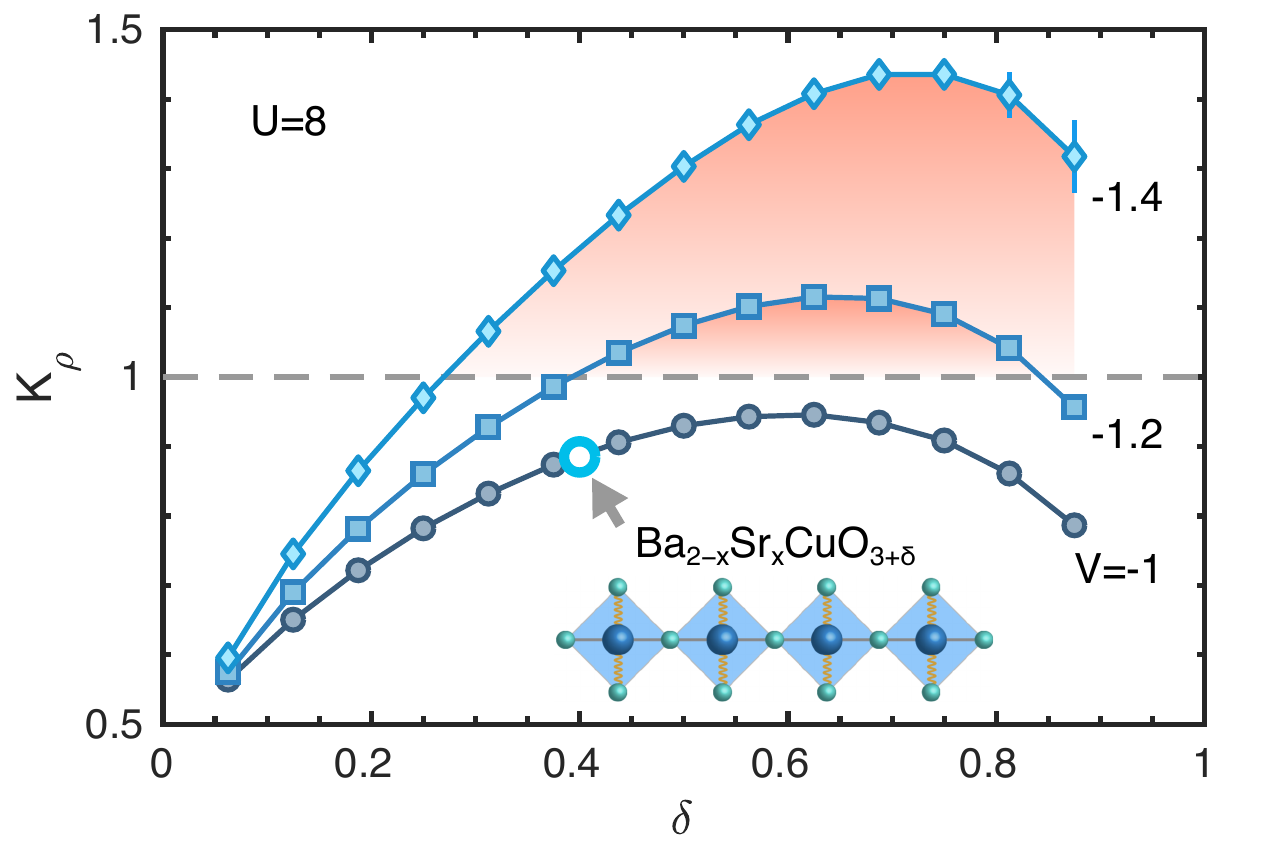}\vspace{-6mm}
\renewcommand{\figurename}{\textbf{Fig.}}
\caption{\textbf{Prediction for possible triplet superconducting phase in 1D 
cuprate chain.}  We show the Luttinger parameter $K_\rho$ versus doping 
$\delta$ for different attractive interactions $V$. In the highlighted regimes 
with $K_\rho>1$, there exists superconductivity phase for sufficiently strong 
attraction $V$ and large doping $\delta$. The blue circle represents the 
cuprate BSCO with the parameters $V=-1$ and $\delta\simeq 40\%$, and
a slight enhancement of the attraction can drive the compound to the triplet 
superconductive phase.}
\label{Fig:Krhovsdelta}
\end{figure}
  
\noindent 
\textbf{TS pairing in the 1D cuprate BSCO.}
Although the phase boundaries, i.e., the critical strengths of $V$, are 
$U$-dependent, they can be determined analytically at quarter filling
in the $U\rightarrow \infty$ limit\,\cite{Luther1975,Schulz1990}. 
In this limit, the Luttinger parameter $K_\rho = 1/[2+(4/\pi) 
\arcsin(v)]$ with $v = V/2$, which exceeds 1 when $|V| \geq \sqrt{2}$. 
According to the recent experiment, the effective model 
parameters for 1D cuprate chain BSCO were identified 
as $U\simeq8$ and $V\simeq-1$\,\cite{Chen2021CuprateChain}. 
Despite anomalously strong, the effective attraction $V$ is still 
slightly below this threshold.

To search for TS in larger parameter space, we further explore the full doping dependence. To approximate the realistic materials, 
we fix $U=8$ and three different values of $V$, and evaluate 
the Luttinger parameter $K_\rho$ for a wide range of doping.
As shown in Fig.~\ref{Fig:Krhovsdelta}, the uniform TS phase 
characterized as $K_\rho>1$ can be realized only if $|V|> 1.2$ 
{(and $|V| \lesssim 1.7$ before PS$_x$ sets in)}, in order to exhibit 
prominent superconducting instability below 40\% doping, the maximal 
accessible doping at current experimental conditions. Therefore, the 
doped BSCO resides on the boundary to a TS phase (as also indicated 
in Fig.~\ref{Fig:PhDiag}\textbf{c}), and a slight reduction of on-site 
$U$ or enhancement of near-neighbor attraction $V$ may drive it 
into the TS phase --- both can be achieved by manipulating the 
electron-phonon coupling either inside the crystal or via a substrate
\cite{Wang2021AttractivePRL}.
\\

\noindent{\bf{Discussion}}\\
Our simulation is based on the recently extracted attractive extended Hubbard 
model for 1D cuprate BSCO from experiments\,\cite{Chen2021CuprateChain}. 
Although this newly demonstrated model and its parameters have been 
theoretically reproduced from the electron-phonon coupling
\cite{Wang2021AttractivePRL}, its impact on emergent phases, 
especially unconventional superconductivity phases, remains 
unknown. In this work, we employ DMRG --- the method of choice 
for 1D correlated systems --- to investigate the EHM with both on-site 
repulsive and near-neighbor attractive interactions. At both half and 
quarter fillings, we identify a prominent gapless TS phase with the 
$p$-wave pairing induced by the attractive interactions. Different 
from the long-range order (hidden) assumption in the context 
of mean-field theory, the $p$-wave superconducting order identified 
in this correlated 1D chain is quasi-long-ranged: the triplet pairing 
correlation $\Phi_{\textrm{T}}$ decays as a power-law at long distance 
and presents as the dominant charge-2e excitations in the gapless TLL, 
and specially, at half filling it dominates over the singlet pairing 
$\Phi_{\textrm{S}}$ also in large distance scaling. Such dominance 
results in divergent triplet superconductive susceptibility at low 
temperature. This phenomenon can be detected by the spectral depletion 
in ARPES or the Drude peak in optical conductivity, both of which 
are accessible for \emph{in situ} synthesized quasi-1D materials. 

As the experimentally extracted model parameters for cuprates
\cite{Chen2021CuprateChain} are close to, though not within, 
the TS phase identified in our simulations, our finding may 
motivate further investigation and manipulation of cuprates 
towards a $p$-wave topological superconductor. Couplings
between the cuprate chains may open a charge gap and introduce 
edge modes that can be very useful in future quantum 
technologies. Due to the chemical and structural similarity
between 1D and 2D  cuprates, our results of the TS phase in 
the attractive EHM here shed light on and call for further
many-body studies of the superconductivity in the EHM of 
higher dimensions
\cite{Senechal2013PRB,Plonka2015PRB,Paki2019PRB,jiang2021arXiv}.

Lastly, our conclusion on the cuprate chain can be extended to other related 
electronic materials. Considering the widely existing electron repulsion and 
electron-phonon coupling, this model with a repulsive $U$ and an attractive 
$V$ may also be applicable, as a low-energy approximation, for other 
transition-metal oxides. There different cuprate compounds and other 
materials may exhibit different microscopic parameters ($U$ and $V$) 
due to their distinct chemical environments, and the rich quantum phases
revealed in the EHM model studies here may find their interesting materialization.
\\

\noindent{\bf{Methods}}\\
\textbf{Density matrix renormalization group.}
We perform DMRG calculations with the charge U(1) and spin 
SU(2) symmetries implemented through the tensor library QSpace
\cite{Weichselbaum2012,Weichselbaum2020}, and compute system 
sizes up to $L=512$ to obtain the spin, charge, and 
superconductive correlations, etc, with high precision. 
In the calculations, we retain up to {$m^*=2048$} multiplets, 
equivalent to $m\approx 4000$ U(1) states, which render 
small truncation errors $\epsilon \lesssim 10^{-7}$. 
We use the open boundary conditions as in conventional 
DMRG calculations. Due to the existence of attraction $V$,
particularly near the PS phase one needs to introduce pinning
fields at both ends and perform sufficient numbers of sweeps 
(even over 100 times) to fully converge the results, e.g., 
the charge distribution along the chain (see Supplementary Note~1).
\\

\section*{Data availability} 
The data that support the findings of this study are 
available from the corresponding author upon reasonable request.\\

\section*{Code availability}
All numerical codes in this paper are available 
upon request to the authors. \\

\section*{Acknowledgements}
We acknowledge Z. Chen, T.P. Devereaux, B. Moritz, 
Z.-X. Shen, Yang Qi, and T. Shi for stimulating discussions. 
W.L. acknowledges the support from the NSFC through 
Grant Nos.~11974036, 11834014, and 12047503. H.C.J. 
was supported by the Department of Energy, Office of 
Science, Basic Energy Sciences, Materials Sciences 
and Engineering Division, under Contract DE-AC02-76SF00515. 
D.W.Q and W.L. thank the High-performance Computing 
Center at ITP-CAS for their technical support and 
generous allocation of CPU time.

\section*{Author contributions}
W.L. and Y.W. initiated this work. 
D.W.Q and B.B.C performed the DMRG calculations.
All authors contributed to the analysis of the results.
W.L., Y.W., and H.C.J. supervised the project.

\section*{Additional information}
\noindent
\textbf{Supplementary Information} is available in the online version of the paper. \\
\noindent
\textbf{Competing interests:} The authors declare no competing interests. \\


%

\newpage\clearpage
\onecolumngrid

\renewcommand{\Fig}[1]{Supplementary Figure~\ref{#1}}
\renewcommand{\Figs}[1]{Supplementary Figures~\ref{#1}}

\begin{center}
{\large Supplementary Information for}
$\,$\\
\textbf{\large{Spin-Triplet Pairing Induced by Near-Neighbor Attraction in the Cuprate Chain}}

$\,$\\
Qu \textit{et al.}
\end{center}



\date{\today}

\setcounter{subsection}{0}
\setcounter{figure}{0}
\setcounter{equation}{0}
\setcounter{table}{0}

\renewcommand{\thesubsection}{\normalsize{Supplementary Note \arabic{subsection}}}
\renewcommand{\theequation}{S\arabic{equation}}
\renewcommand{\thefigure}{\arabic{figure}}
\renewcommand{\thetable}{\arabic{table}}

\subsection{DMRG Techniques in Simulating the Extended Hubbard Model}
\label{SN:DMRG}

\textbf{Relieving the boundary effects by pinning fields.}
At half filling and especially near the PS$_2$ phase boundary,
there is strong boundary effects in the {DMRG} calculations, 
which significantly lower the simulation efficiency. 
To relieve this problem and accelerate the calculations, 
we apply a pinning field term $-V(n_i-1)^2$ to both boundaries
$i=1,L$, where $n_i=n_{i\uparrow}+n_{i\downarrow}$ is the 
local particle number operator at site $i$. Here we set the 
strength of the pinning term proportional to $V$, so that 
the boundary effect can be well reduced for various 
cases with different attraction strengths.

\begin{figure}[!btp]
\includegraphics[width=1\linewidth]{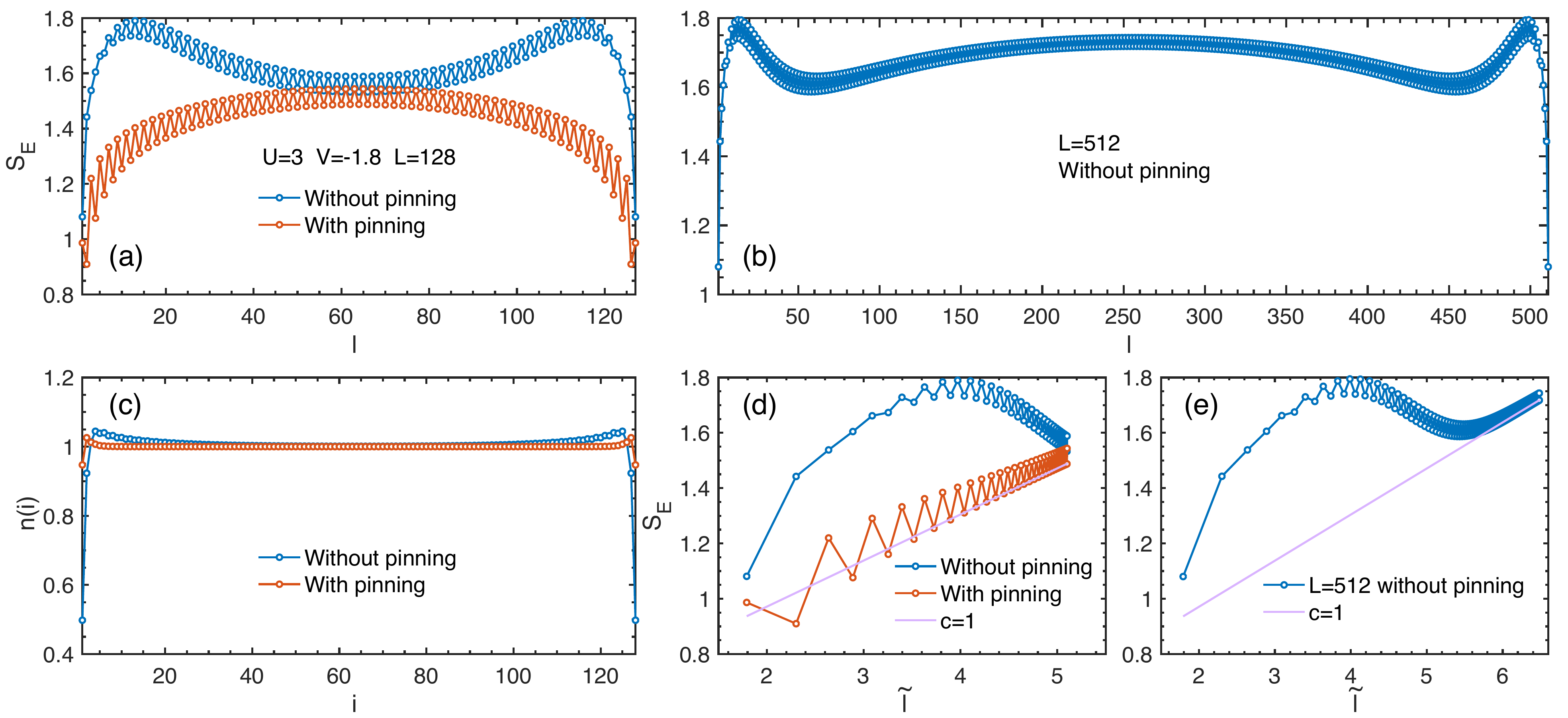}
\renewcommand{\figurename}{\textbf{Supplementary Figure}}
\caption{
\textbf{DMRG results of the EHM for $U=3$ and $V=-1.8$ with and without pinning fields.}
The entanglement entropy results $S_\textrm{E}(l)$ are shown along 
the (a) $L=128$ chain (with and without pinning fields) 
and (b) $L=512$ without pinning. (c) Real-space charge 
density distribution on $L=128$ chain with and without 
pinning. (d,e) show the same $S_\textrm{E}$ data as in (a,b),
while plotted versus the conformal distance $\tilde{l}$. 
In (d) the data with pinning clearly exhibits a logarithmic
scaling corresponding to a central charge of $c=1$. In 
panel (e), only $S_\textrm{E}$ data deep in the bulk shows the 
expected $c=1$ behavior.
}
\label{Fig:Pinning}
\end{figure}

Such pinning terms do help the DMRG to relieve the strong
boundary effects and converge to the uniform charge 
distribution in the bulk. For example, in \Fig{Fig:Pinning} 
we show the results of half-filled EHM at $U=3$ and 
$V=-1.8$ (i.e., in SDW phase) with and without pinning. 
There, the entanglement entropy is defined as 
$$S_\textrm{E}(l)= -\operatorname{Tr}
[\rho(l)\ln \rho(l)],$$ where $\rho(l)$ is the reduced 
density matrix of a subsystem with length $l$. In
\Fig{Fig:Pinning}(a), we see although the $S_\textrm{E}(l)$ 
distribution is strongly affected by boundary effects without 
pinning, such modulation can be ``healed" by the pinning terms. 
An alternative way to control the boundary effects is to push 
the calculations to longer system sizes. In \Fig{Fig:Pinning}(b),
we compute a very large system with $L=512$, where
$S_\textrm{E}(l)$ shows the expected dome shape deep 
in the bulk, despite that the boundary effect still 
penetrates into the bulk with over 50 sites.

According to the conformal field theory
\cite{Calabrese2004,Fagotti2011}, for 1+1 dimensional 
critical systems, the entanglement $S_\textrm{E}$ should 
follow a linear scaling with the conformal distance 
$$S_\textrm{E}=\frac{c}{6} \cdot \tilde{l} + \rm{const.},$$ 
where $\tilde{l} =\ln\left[ \frac{4(L+1)}{\pi}\sin
\frac{\pi(2l+1)}{2(L+1)}\right]$ with $L$ the total system 
size. In the SDW phase with only one gapless spin mode, we 
have central charge $c=1$ that can be obtained by fitting 
the $S_\textrm{E}$ data. However, a naive analysis fails for the 
$L=128$ case due to strong finite-size effects as shown by 
the blue symbol and lines in \Fig{Fig:Pinning}(d). When the
pinning term is turned on, the $S_\textrm{E}(l)$ curve 
exhibits a usual dome shape and is consistent with central 
charge of $c=1$ even for a moderate system size $L=128$, 
as shown in \Fig{Fig:Pinning}(a) and (d). Moreover, 
the pinning field makes the charge density distribution 
$n(i)$ more homogeneous (thus closer to half-filling 
throughout the chain), as shown in \Fig{Fig:Pinning}(c). 
By pushing the calculations to larger sizes, like $L=512$ in
\Fig{Fig:Pinning}(e) we plot $S_\textrm{E}$ vs. $\tilde{l}$
and find it falls into a linear relation in the bulk (large
$\tilde{l}$ regime), also confirming the central charge $c=1$
in the SDW phase. 

In conclusion, the pinning fields on the boundaries of the
attractive-$V$ EHM can significantly reduce the boundary effects,
making the results much more accurate and well-behaved for
analysis. Fortunately, at other fillings the problem is less
severe, and it is also quite hard to find a proper pinning 
strength for every $U$, $V$ and doping. Therefore in the
quarter-filling calculations we did not employ the trick.
\\

\begin{figure}[!tbp]
\includegraphics[width=1\linewidth]{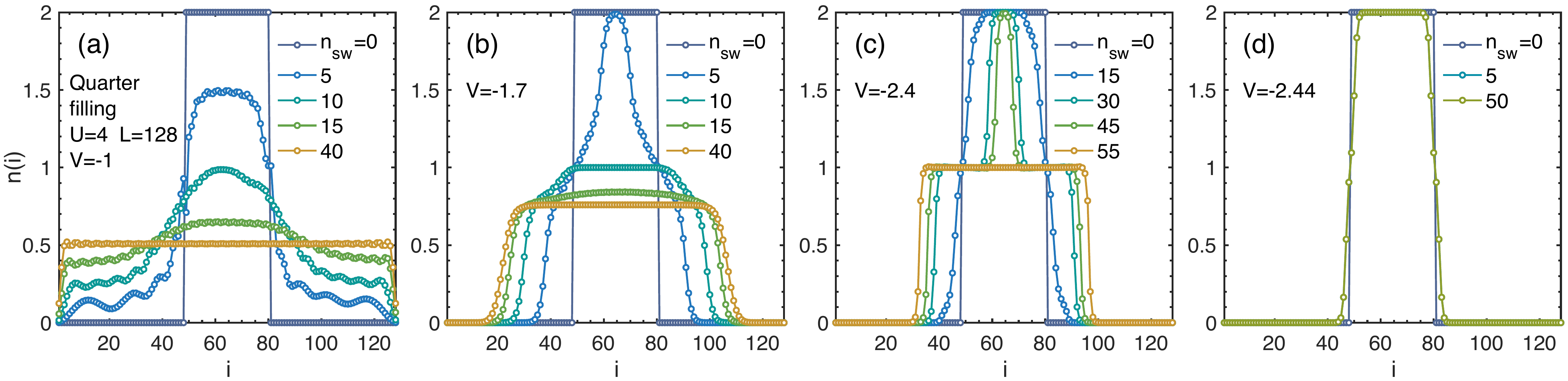}
\renewcommand{\figurename}{\textbf{Supplementary Figure}}
\caption{
\textbf{The charge density distribution $n(i)$ of EHM vs. DMRG 
sweeps $n_\text{sw}$.} (a-d) show the evolution of $n(i)$ with 
$n_\text{sw}$ up to 55, for various interactions $V=-1,-1.7,-2.4,-2.44$ 
(and with $U=4$, $L=128$), respectively. The converged distribution 
$n(i)$ are those shown in Fig.~5\textbf{c} in the main text. 
}
\label{Fig:CDenEvol}
\end{figure}

\textbf{Wave function initialization and DMRG sweeps.}
Since there are multiple phase separation phases and 
first-order phase transitions in the phase diagram
Fig.~1 of the main text, it is good 
to start with a proper initial wave function for later 
DMRG sweeps, so as to avoid being trapped in local 
minima. Through careful numerical tests, we found the 
so-called PS$_2$ initialization constitutes a very good 
option in our practical calculations, where the particle 
number is doubly occupied ($n=2$) in the middle and $n=0$
towards the ends (c.f., the $n_\text{sw}=0$ lines in
\Fig{Fig:CDenEvol}). 

In Fig.~5\textbf{c} of the main text we have shown the
charge density distribution $n(i)$ at quarter filling,
in various phases of the EHM. Specifically, in \Fig{Fig:CDenEvol}
we show how the initial PS$_2$ state evolves vs. DMRG sweeps
$n_{\rm sw}$, which finally converges to the distribution 
that corresponds to the true ground state. For example,
in \Fig{Fig:CDenEvol}(a) we find, for $V=-1$ in the TS phase,
the initially clustered part --- the $n=2$ plateau --- gradually
smears out and expands to the whole system, recovering 
finally a virtually uniform distribution. For $V=-1.7$ 
in the PS$_x$ phase, the $n=2$ plateau firstly shrinks 
into a $n=1$ one and then to the final $n=x$ plateau in 
the central of the chain, as shown in \Fig{Fig:CDenEvol}(b). 
For $V=-2.4$ in the PS$_1$ phase, as shown in 
\Fig{Fig:CDenEvol}(c), an $n=1$ plateau appears during 
the sweeps, which eventually occupies the whole electron 
cluster in the PS$_1$ state. Finally, in 
\Fig{Fig:CDenEvol}(d), the distribution remains the 
same shape as $n_{\rm sw}$ increases in the case of $V=-2.44$
(PS$_2$ phase). To summarize, the PS$_2$ wave function 
works well as the initial state for simulating various 
phases in our practical calculations.
\\

\begin{figure}[!tbp]
\includegraphics[width=0.8\linewidth]{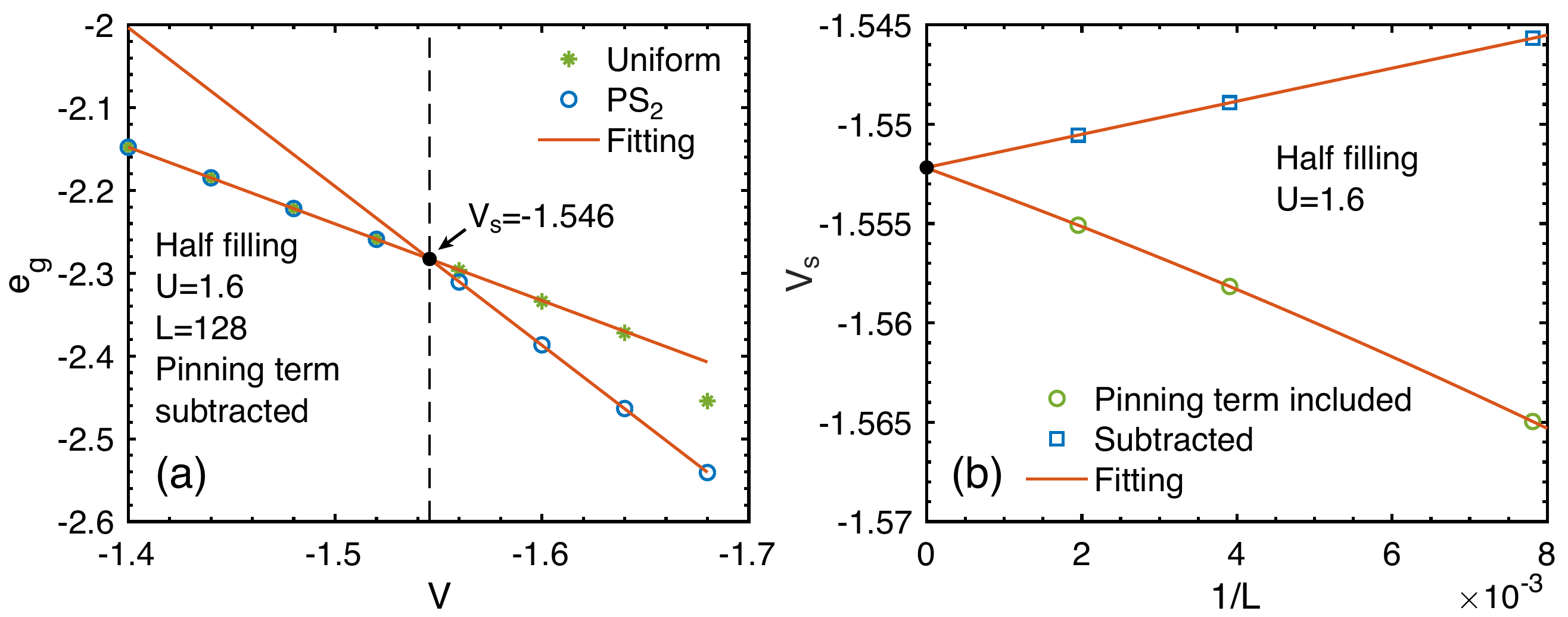}
\renewcommand{\figurename}{\textbf{Supplementary Figure}}
\caption{
\textbf{Determination of the TS-PS$_2$ transition point $V_\textrm{s}$.} 
(a) shows the ground-state energy $e_\textrm{g}$ (per site, 
labeled by the open circles and asterisks) obtained with 
two different initial states, namely, the charge uniformly 
distributed state and the PS$_2$ initialization. The pinning 
term contribution has been subtracted from $e_\textrm{g}$, and the two 
red lines are linear fittings of the $e_\textrm{g}$ data from opposite 
sides of the transition point (circles from the right and 
asterisks from the left). The crossing point provides an 
accurate estimation of $V_\textrm{s}$. (b) shows the estimated $V_\textrm{s}$ 
from two schemes: including or subtracting the pinning term
contributions. The red lines are second-order polynomial 
fittings, and the extrapolations to $1/L=0$ provide a highly 
consistent estimate of the transition point $V_\textrm{s}$.
}
\label{Fig:DetermineVPS2}
\end{figure}

\textbf{Determination of the PS$_2$ Boundary.}
Here we show the details of determination of condensation
transition boundary in the half-filling phase diagram in
Fig.~1\textbf{b} of the main text. As discussed above, 
the PS$_2$ initialization works very well in practice.
Therefore, with this initial state we perform the DMRG
calculations of different $V$ values, and obtain an 
estimation of the transition point $V_\textrm{s}$ for fixed $U$.
However, when $V$ is close to the condensation transition 
point $V_\textrm{s}$, the convergence becomes quite slow. 
To determine $V_\textrm{s}$ more efficiently, we compare the 
energy $e_\textrm{g}$ {(per site)} versus $V$ computed with two 
different initial states, i.e., the PS$_2$ and a uniform 
initial state. For example, in \Fig{Fig:DetermineVPS2}(a) 
we show the energy $e_\textrm{g}$ versus $V$, obtained with the 
two different initial states. We find in the TS phase 
the two $e_\textrm{g}$ values obtained with different initial 
states are in excellent agreement, while 
in the PS$_2$ phase they are different: the PS$_2$ 
initial state leads to a significantly lower energy. 
Therefore we extrapolate the energy curve from two sides 
of the transition through a linear fitting, and find the
crossing point [the black dot in 
\Fig{Fig:DetermineVPS2}(a)]
constitutes an accurate estimation of $V_\textrm{s}$. 

In \Fig{Fig:DetermineVPS2}(a) we have subtracted 
the contribution of the pinning term in the energy $e_\textrm{g}$,
which, nevertheless, can be remained in the analysis.
In \Fig{Fig:DetermineVPS2}(b) we plot the estimated $V_\textrm{s}$ 
from different system sizes, and compare the two schemes, 
i.e., including or excluding the pinning term contributions 
in $e_\textrm{g}$. We see the results of the two schemes approach 
each other as $L$ is increased. By a second-order polynomial
fitting and extrapolation to $L=\infty$, the two schemes 
give a consistent result. However, the $V_\textrm{s}$ estimated by
subtracting the pinning term are found to converge faster, 
and in practice the $L=128$ result is already precise enough 
with a relative error only about $10^{-3}$. Therefore, 
in the practical calculations we stick to such a fast 
scheme and use the $L=128$ data to obtain an accurate 
estimation of $V_\textrm{s}$.
\\

\subsection{Spin, Charge, and Pairing Correlations in the 
Tomonaga-Luttinger Liquid Theory}
\label{SN:TLL}

The Tomonaga-Luttinger liquid (TLL) theory can be used 
to describe a large family of one-dimensional (1D) quantum 
critical states~\cite{Haldane1981Luttinger,Voit_RepProgPhys1995,
Giamarchi1D,FradkinFieldTheory}. Here we briefly recapitulate
certain results from the TLL theory, relevant to the current 
work, on the spin, charge, and pairing correlations used in 
the analysis of our DMRG results in the main text. For the
results listed below, we focus on the cases with spin SU(2)
symmetry and have the Luttinger parameter $K_\sigma=1$ for 
the spin gapless states.

In general, there exist multiple modes for a given two-point
correlation functions. For example, consider the charge 
density-density correlation $D(r) = \langle n_i n_j \rangle 
- \langle n_i \rangle \langle n_j \rangle$ with $n_{i(j)}$ 
the particle number operator at site $i(j)$ and 
$r\equiv j-i$, and the spin correlation $F(r) = 
\langle \mathbf{S}_{i} \cdot \mathbf{S}_{j} \rangle$
with $\mathbf{S}_{i(j)}$ the spin operator at site $i(j)$.
Up to the first two dominant modes, both correlations 
have a similar form as
\cite{Giamarchi1D,Schulz1990,Kawakami1990}
\begin{eqnarray}
 D(r) = -\frac{K_\rho}{(\pi r)^2} + A_1 \frac{\cos(2k_F r)}{r^{1+K_\rho}} \ln^{-3/2}(r),\\
 F(r) = -\frac{3}{4}\frac{1}{(\pi r)^2} + B_1 \frac{\cos(2k_F r)}{r^{1+K_\rho}} \ln^{1/2}(r),
\end{eqnarray}
where $A_1$ and $B_1$ are model-dependent parameters.
Generally, when $K_\rho<1$ the $2k_F$ oscillation term 
dominates in both the charge and spin correlations, 
while for $K_\rho>1$ the uniform $r^{-2}$ term becomes 
the leading one. These different modes are reflected in 
the singularities of the structure factor $S(k)$. 
The uniform mode results in 
$S(k)\simeq \tilde{K}_\nu |k|/\pi$ for $k\to 0$, where $\tilde{K}_{\nu=\rho}
=K_\rho$ for charge and $\tilde{K}_{\nu=\sigma}=\frac{3}{4}K_\sigma$ for spin; 
while for the $2k_F$ mode, it leads to $S(k)\sim c_1 + c_2|k\mp 2k_F|^{K_\rho}$ for $k\to \pm 2k_F$.

Next we consider the two superconductive pairing correlations.
The singlet pairing correlation function is defined as
$\Phi_\textrm{S} (r) = \langle \Delta_\textrm{S}^\dag (i) \Delta_\textrm{S} (j) \rangle$, 
where $\Delta^\dag_\textrm{S}(i) = \frac{1}{\sqrt{2}} (c^\dag_{i,\uparrow}
c^\dag_{i+1,\downarrow} - c^\dag_{i,\downarrow} c^\dag_{i+1,\uparrow} )$, 
whose two leading algebraic modes take the scaling form
\cite{Kawakami1990,Pruschke1992}
\begin{equation}\label{Eq:PhiSScaling}
\Phi_\textrm{S}(r) = \frac{C_0}{r^{1+1/K_\rho}} + C_1\frac{\cos(2k_F r)}{r^{K_\rho+1/K_\rho}},
\end{equation}
where the logarithmic correction is neglected. Therefore, 
for $K_\rho<1$ the $2k_F$ term dominates, while for $K_\rho>1$ 
the uniform term takes over and the overall pairing correlation
decays algebraically with an exponent $1+1/K_\rho<2$. On the 
other hand, the (averaged) triplet pairing correlation 
$\Phi_\textrm{T}(r) = \frac{1}{3} \sum_{s=\pm1,0} \Phi_{\textrm{T},s} (r)$ 
scales as
\begin{equation}
\Phi_\textrm{T}(r) = \frac{D_0}{r^{1+1/K_\rho}} + D_1\frac{\cos(2k_F r)}{r^{K_\rho+1/K_\rho+2}},
\end{equation}
with the logarithmic corrections also omitted. Different 
from $\Phi_\textrm{S}$, here the $2k_F$ term is much weaker and the 
uniform term is always playing a dominant role in long 
distance $r$. This is indeed what we have observed in our 
DMRG calculations of $\Phi_\textrm{T}$, as can be seen in
Fig.~3 of the main text as well as
\Figs{Fig:SupportQF} and \ref{Fig:PSxSC} below.

In addition, we note the coefficients $A_1$, $B_1$, $C_0$, 
$C_1$, etc.~are model-dependent, and can take very different 
values or even be absent. For example, in the main text we have 
found $\Phi_\textrm{S}$ only has $2k_F$ mode 
in the half-filled TS phase, i.e. $C_0\simeq0$, based on our
accurate DMRG calculations.

Besides the above charge-2e correlations, the single-particle
Green's function $G(r)= \sum_\sigma \langle c_{i\sigma}^\dag
c_{j\sigma} \rangle$ also shows a power-law behavior as
$r^{-1-\alpha_G}$ with $\alpha_G = 2\sum_\nu
\gamma_\nu$~\cite{Voit1993}, where 
\begin{equation}
\label{EqS:GF}
\gamma_\nu = \frac{1}{8} \left( K_\nu + \frac{1}{K_\nu} -2 \right), \quad
\nu=\sigma,\rho.
\end{equation}
As mentioned above, for the EHM with SU(2) spin symmetry, 
we always have $K_\sigma=1$. Therefore, $G(r)$ has the slowest
decaying power $-1$ when $K_\rho=1$ (at the SDW-TS transition,
TLL-TS crossover, etc).

By taking a Fourier transformation of $G(r)$, we obtain the 
occupation number distribution $n(k)$ in momentum space. 
In the TLL phase, although in general $n(k)$ is continuous 
at $\pm k_F$ (except for the $K_\rho=1$ case), there is 
nevertheless singularity in the form of 
$n(k)\sim c_3 + c_4|k\pm k_F|^{\alpha_G}{\rm sign} 
(k\pm k_F)$ right at the Fermi vector $k_F$.
\\

\subsection{Determination of the Luttinger Parameter $K_\rho$}
\label{SN:DetermineKrho}

\begin{figure}[!tbp]
\includegraphics[width=1\linewidth]{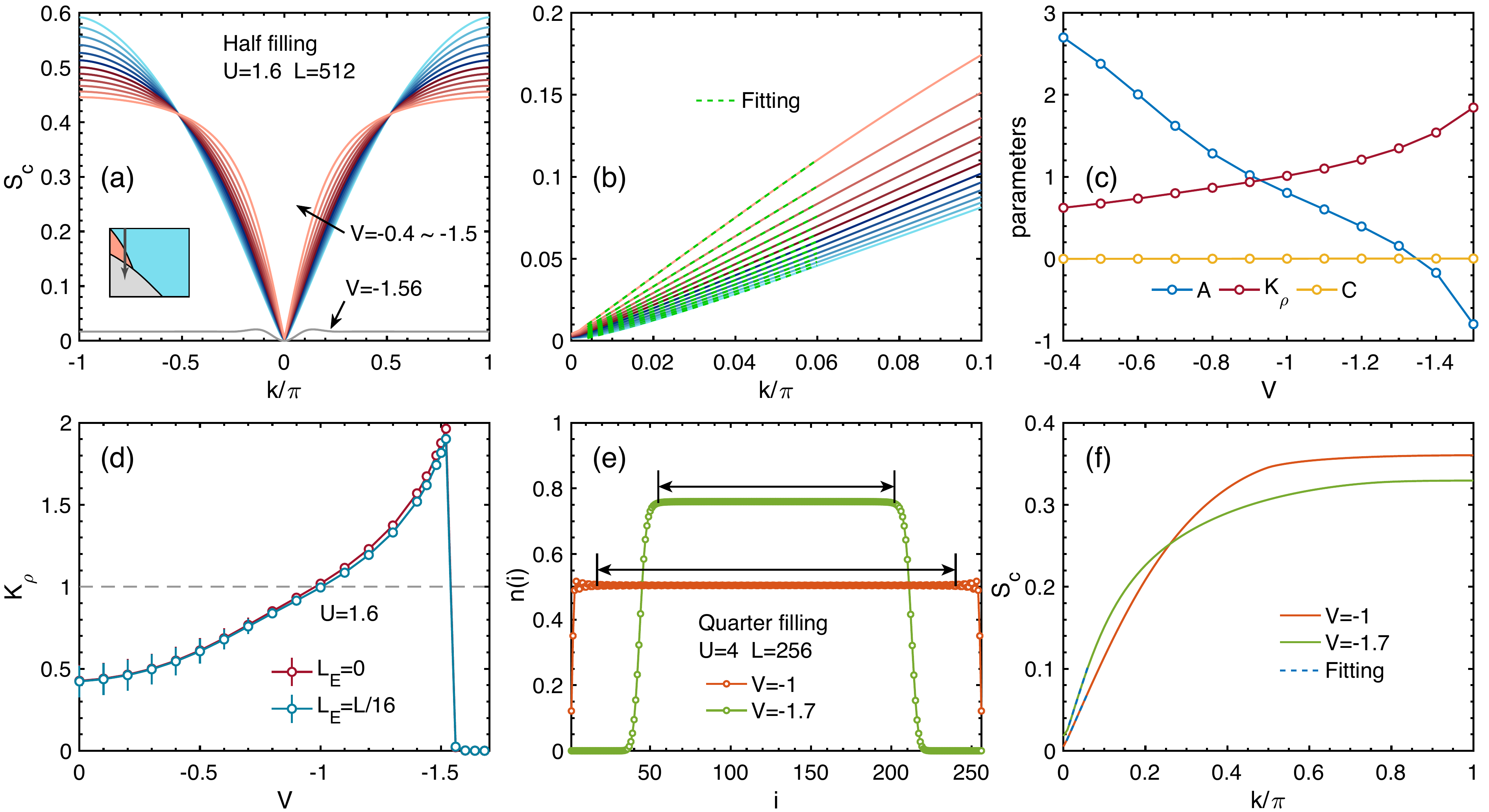}
\renewcommand{\figurename}{\textbf{Supplementary Figure}}
\caption{
\textbf{Determination of the Luttinger parameter $K_\rho$.} 
{Here we show the results of} EHM at (a-d) half filling and (e,f) quarter 
filling. (a) shows the charge structure factor $S_\textrm{c}(k)$ 
at various $V$, with the small $k$ regime zoomed in and 
shown in (b). The dashed green lines in (b) are the second-order polynomial fittings 
$S_\textrm{c}(\tilde{k}) = A \tilde{k}^2 + K_\rho \tilde{k} + C$, 
where $\tilde{k}=k/\pi$. The fitted parameters $A$, $K_\rho$,
and $C$ are shown in panel (c) as functions of $V$. In panel
(d) we compare the results of $L_E=0$ (no sites skipped) and
$L_E=L/16$ (with $L/16$ sites skipped from both ends). The
error bar represents difference in taking wider or narrower 
$k$ regimes to fit. (e) shows the charge density distribution
$n(i)$ at quarter filling, and the double-arrows indicate the
clustered part used in computing related $S_\textrm{c}(k)$, with the
results (solid line) and linear fittings (dashed) $S_\textrm{c}(\tilde{k})=
K_\rho \tilde{k} + B$ shown in panel (f).
}
\label{Fig:DetermineKrho}
\end{figure}

As discussed in \ref{SN:TLL}, the Luttinger 
parameter $K_\rho$ can be extracted from the charge 
structure factor, i.e., $S_\textrm{c}(k) \simeq K_\rho |k|/\pi$ 
for $k\rightarrow 0$. Here $S_\textrm{c}(k)$ is the Fourier transformation (FT) of the charge correlation function
\begin{equation}
\label{Eq:Sc}
S_\textrm{c}(k) = \frac{1}{L}\sum_{i,j} e^{\mathrm{i} k (i-j)} D(i,j), 
\end{equation}
where $D(i,j) = \langle n_i n_j \rangle - \langle n_i \rangle 
\langle n_j \rangle$. 
In the standard FT, the momentum $k$ takes discrete value $0$,
$\frac{2\pi}{L}$, $\frac{4\pi}{L}$, ... , $\frac{2\pi(L-1)}{L}$. 
To collect more momentum data points for the purpose of fitting,
we extend the value of $k$ to the continuum in $[0,2\pi]$ 
(or equivalently $[-\pi, \pi]$). That is, we still compute
$S_\textrm{c}(k)$ using \Eq{Eq:Sc} but now the $k$ can take 
the continuum of values in the Brillouin zone, which constitutes
a natural and smooth interpolation method.

For the charge-gapless phase like TLL, the asymptotic 
behavior of $S_\textrm{c}(k)$ approaching $k=0$ is linear 
with $k$, while in the charge-gapped phase like the SDW we 
have $K_\rho=0$ and the small-$k$ scaling is instead 
quadratic. This offers us an efficient way to extract $K_\rho$ 
from the charge structure factor $S_\textrm{c}(k)$. In practice, 
to reduce the boundary effects, when we compute $S_\textrm{c}(k)$ 
the left- and right-most $L_E$ edge points are skipped 
and only the bulk correlation data are used. By doing so, 
we observe that the the determined $K_\rho$ results are 
not very sensitive to $L_E$ and thus numerically accurate, 
as shown \Fig{Fig:DetermineKrho}(d). 

At half filling, as the SDW and TS phases are respectively
charge gapped and gapless, we employ the second-order 
polynomial fitting $S_\textrm{c}(\tilde{k}) = A \tilde{k}^2 + 
K_\rho \tilde{k} + C$, where $\tilde{k}=k/\pi$ is the
(normalized) momentum. The intercept $C$ is introduced 
as the discarded edge sites can break the particle number
conservation, making $S_\textrm{c}(0)\neq 0$. The $S_\textrm{c}(k)$ results 
for $U=1.6$ at half filling are shown in
\Fig{Fig:DetermineKrho}(a). As the charge gap in the SDW 
phase is small in the weak coupling regime of $U=1.6$, 
the change of $S_\textrm{c}(k)$ from quadratic to linear behaviors 
can be seen clearly if we zoom in into the small $k$ 
regime in \Fig{Fig:DetermineKrho}(b). Since the quadratic 
region in SDW phase is narrow, we need to carefully choose 
the momentum range $[k_\text{min}, k_\text{max}]$ for our
fittings. We set the lower bound of the fitting range to
$k_\text{min} =  2\pi/L$ (to avoid finite-size gap effects),
and different upper bounds $k_\text{max}$ ranging from
$0.03\pi$ to $0.06\pi$ have also been chosen in practical
fitting. We take average of the fitting results to provide 
a reliable estimate of $K_\rho$ with error bars. As shown 
in \Fig{Fig:DetermineKrho}(c), indeed we observed that 
in the SDW phase the quadratic coefficient $A$ is predominant 
while in the TS phase ($V<V_\textrm{c}\simeq -1$) the linear 
coefficient $K_\rho$ dominates.

At quarter filling the charge sector is always gapless 
in the uniform TLL and TS phases, and thus we can employ 
a linear fitting $S_\textrm{c}(k)=K_\rho \tilde{k} + B$ throughout.
Besides, we are also interested in the clustered regime 
in phase separation phase PS$_x$ and extract the Luttinger
parameter $K_\rho$ of the clustered electrons. To obtain that,
we select the bulk correlations measured on the central regime,
with the criteria that the local particle density being very
close to and at lease $99\%$ of the density in the very center
[c.f., \Fig{Fig:DetermineKrho}(e)]. We show the computed
$S_\textrm{c}(k)$ data and their linear fittings in
\Fig{Fig:DetermineKrho}(f), with the $K_\rho$ results in
Fig.~5\textbf{a} of the main text.
\\

\subsection{Fitting the Central Charge $c$}
\label{SN:CentralCharge}

\begin{figure}[!tbp]
\includegraphics[width=1\linewidth]{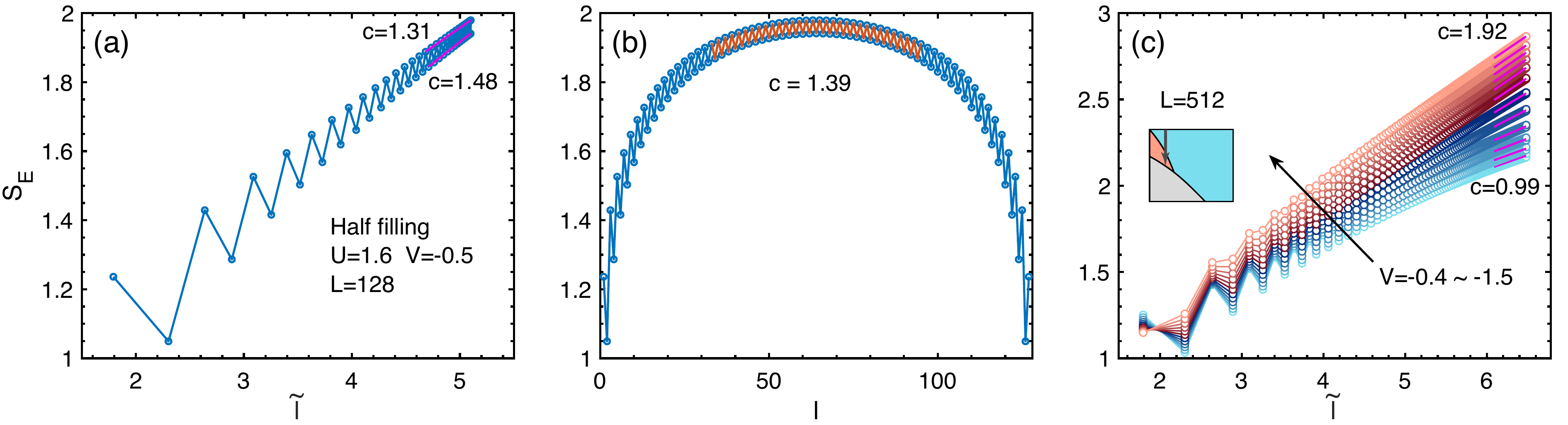}
\renewcommand{\figurename}{\textbf{Supplementary Figure}}
\caption{\textbf{Fitting the central charge $c$ from the 
entanglement entropy results $S_\textrm{E}$ of the half-filled EHM.}
(a) shows the $S_\textrm{E}$ result of the EHM and its direct 
linear fittings with Eq.~(\ref{EqS:SE}), which gives 
slightly different $c$ for the up and down branches. 
(b) The same $S_\textrm{E}$ data as (a) while analyzed using
entanglement scaling formula Eq.~(\ref{EqS:SE2}). (c)
Entanglement data computed on a large system $L=512$ 
are fitted with Eq.~(\ref{EqS:SE2}), and we plot them 
vs. $\tilde{l}$ by the pink lines for various 
interactions $V$.
}
\label{Fig:CentralCharge}
\end{figure}

For the 1+1 dimensional quantum critical states described 
by the conformal field theory~\cite{Calabrese2004,Fagotti2011}, 
the central charge $c$ constitutes an important characteristic
of the universality class of quantum criticality, which can be 
extracted by fitting the entanglement data by the scaling form
\begin{equation}
\label{EqS:SE}
S_\textrm{E}=\frac{c}{6} \, \tilde{l} + \rm{const.},
\end{equation}
where 
$$\tilde{l} = \ln\left[ \frac{4(L+1)}{\pi}\sin\frac{\pi(2l+1)}{2(L+1)}\right]$$ 
is the conformal distance.

In practical calculations with the open boundary condition
(OBC), we observe oscillations in $S_\textrm{E}$ that disturb the perfect
scaling in Eq.~(\ref{EqS:SE}), especially in the SDW phase. 
In \Fig{Fig:CentralCharge}(a) we show the $S_\textrm{E}$ results of 
the half-filled EHM with $U=1.6$ and $V=-0.5$. There 
is apparent period-two oscillation that splits the $S_\textrm{E}$
curve into two branches, and by fitting each of them we get 
slightly different estimates of the central charge $c$,
as shown in \Fig{Fig:CentralCharge}(a). To relieve 
this problem, we use the OBC scaling form of 
$S_\textrm{E}$~\cite{Fagotti2011}
\begin{equation}\label{EqS:SE2}
S_\textrm{E}(l) = \frac{c}{6} \ln\left[\frac{4(L+1)}{\pi}\sin\frac{\pi(2l+1)}{2(L+1)} 
|\sin k_F| \right] + b \frac{\sin[k_F (2l+1)]}{\frac{4(L+1)}
{\pi}\sin\frac{\pi(2l+1)}{2(L+1)}|\sin k_F|} + a.
\end{equation}
To further reduce the boundary effects, we use the bulk 
$L/2$ sites for fitting. In \Fig{Fig:CentralCharge}(b) 
we fit the same data in \Fig{Fig:CentralCharge}(a) with 
this formula, which gives improved fittings and a better estimation of the central charge $c$.

With this improved fitting formula for OBC data, 
in \Fig{Fig:CentralCharge}(c) we analyze the entanglement 
data on a large system $L=512$, from which it can be found
that as $|V|$ increases the fitted central charge $c$ 
changes from $c=1$ to $c=2$, which are collected and 
plotted in Fig.~2\textbf{b1} of the main text.
\\

\subsection{More DMRG Results at Half Filling}
\label{SN:HF}

\begin{figure}[!tbp]
\includegraphics[width=.55\linewidth]{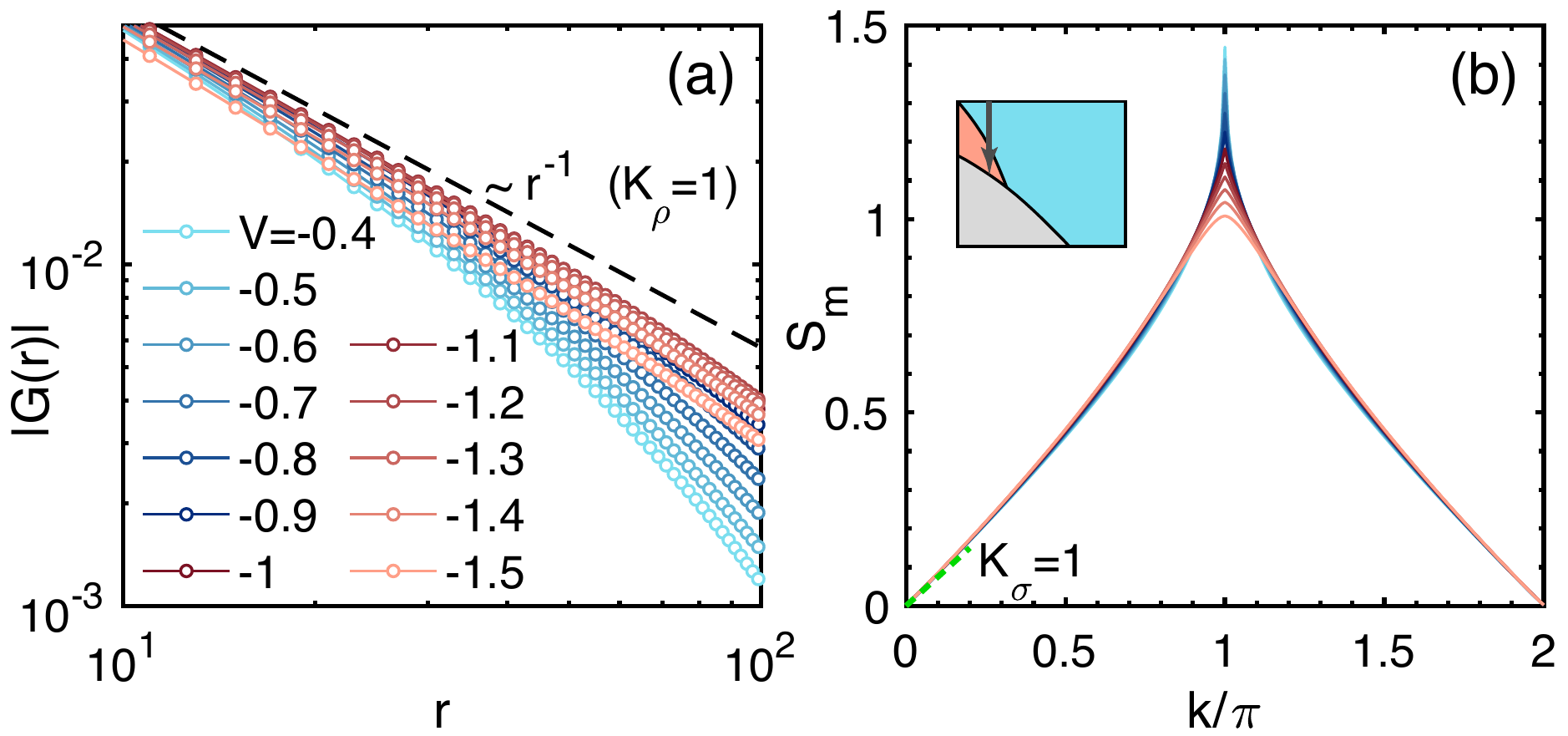}
\renewcommand{\figurename}{\textbf{Supplementary Figure}}
\caption{\textbf{Green's function and spin structure factor results 
of the EHM with $U=1.6$ and at half filling.} (a) Single-particle Green's function $G(r)$ for various 
interactions $V$, with data of only odd distances plotted. 
The dashed line represents the power-law behavior $~r^{-1}$
that corresponds to $K_\rho=1$. (b) The spin structure factors
with various attractions $V$, the $S_\textrm{m}$ curves overlap with 
each other near $k=0$ (and $2\pi$) and can be well fitted by
the dashed green line corresponding to $K_\sigma=1$. A divergent
peak in the SDW phase with relatively weak attraction $V$
becomes a cusp in the intermediate-$V$ TS phase.
}
\label{Fig:SupportU1.6}
\end{figure}

Here we provide more supportive data in the determination 
of quantum phase diagram of the EHM at half filling.\\

\textbf{The $U=1.6$ case.}
As discussed in \ref{SN:TLL}, the single-particle 
Green's function $G(r)$ decays slowest as $r^{-1}$ when 
$K_\rho=1$. Indeed, in \Fig{Fig:SupportU1.6}(a) we see $G(r)$ 
decay exponentially for $V\gtrsim V_\textrm{c}\simeq-1$ in the SDW 
phase but approaches the upper bound $~r^{-1}$ (the dashed 
line) as the attraction $V$ strengthens. Then for 
$V\lesssim V_\textrm{c} \simeq -1$ in the TS phase, it decays 
in power law and departs from the $~r^{-1}$ scaling as 
$K_\rho>1$.  

Regarding the spin correlations, although there is always
$2k_F=\pi$ singularity in the spin correlations in both the 
SDW and TS phases, there are still clear distinctions. 
In \Fig{Fig:SupportU1.6}(b) we see there is divergent 
behavior in the spin structure factor $S_\textrm{m}(k)$ at
$k=\pi$ for $V\gtrsim V_\textrm{c}$. This is because the long-range
spin correlation in the SDW phase scales as $(-1)^r/r$, 
and $S_\textrm{m}(\pi)$ thus diverges as $\ln(L)$ in the
thermodynamic limit. However, within the TS phase the 
exponent of the $2k_F=\pi$ correlation is $1+K_\rho>2$, 
meaning the spin correlation decays faster than $1/r$ 
and corresponds to a non-divergent $S_\textrm{m}(\pi)$. 
In \Fig{Fig:SupportU1.6}(b) we indeed observe these 
features, and in particular find there only a cusp in
$S_\textrm{m}(\pi)$ in TS phase with $V\lesssim V_\textrm{c}$.
\\

\begin{figure}[!tbp]
\includegraphics[width=1\linewidth]{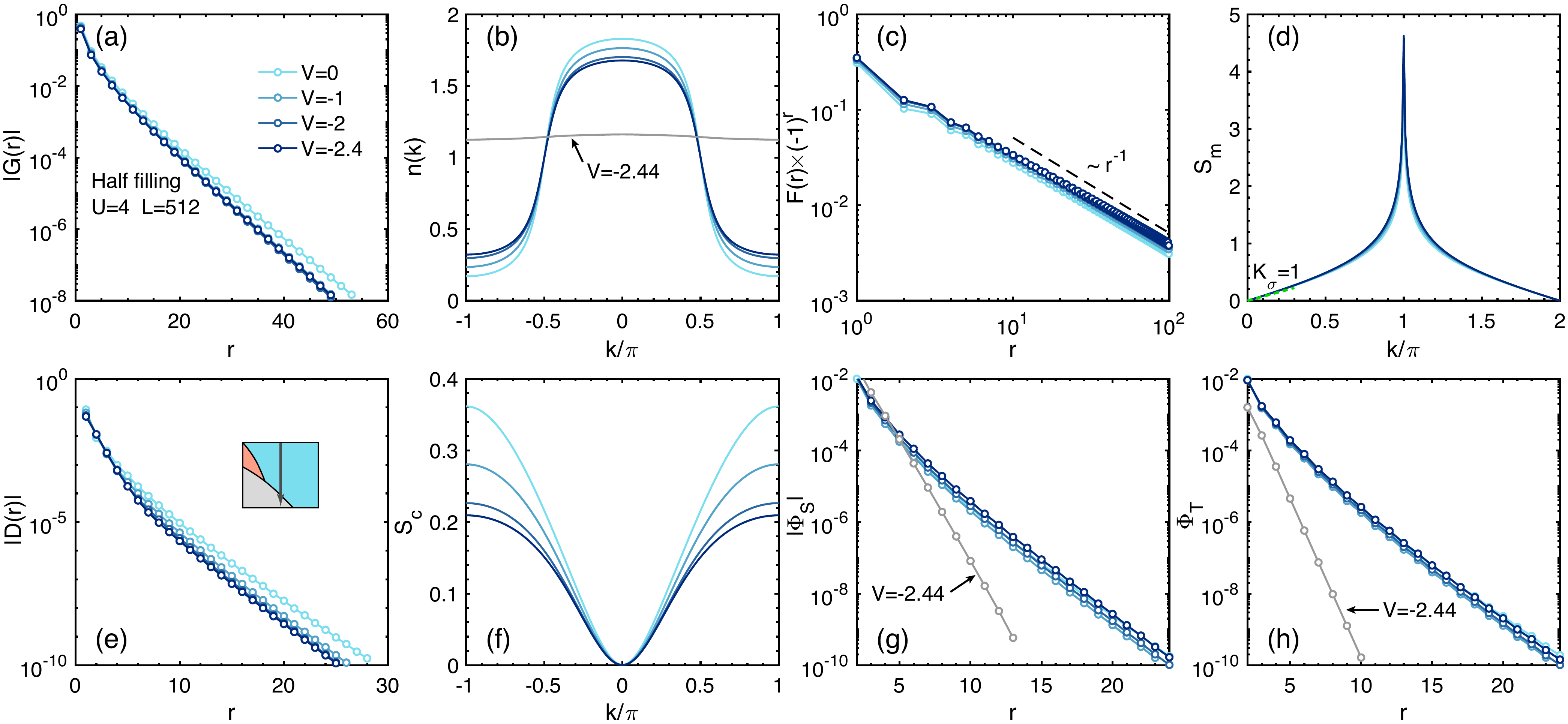}
\renewcommand{\figurename}{\textbf{Supplementary Figure}}
\caption{\textbf{Various correlation functions and structure 
factors of the EHM at half filling and with a fixed 
$U=4$.} The interaction $V$ is scanned from $0$ to 
$-2.4$ in all eight panels with the same legend 
in (a), except for the $V=-2.44$ data that has 
been specially annotated. (a) The single-particle 
Green's functions $G(r)$ show exponential decay 
for various interactions $V$ (only the data for 
odd distances are plotted). (b) Particle number 
distribution $n(k)$ in momentum 
space for various $V$, where no singularity at 
$k = \pm k_F = \pm \pi/4$ is observed. The curve 
for $V=-2.44$ in the PS$_2$ phase is almost flat.
(c) The spin correlation functions $F(r)$ in the 
SDW phase virtually coincide for various $V$, 
and all show $~(-1)^r/r$ power-law scaling. 
(d) The spin structure factors $S_\textrm{m}(k)$ exhibit 
sharp diverging peak at $k=\pi$ for various 
interactions $V$. The dashed green line denotes 
the linear relation reflecting $K_\sigma=1$.
(e) The charge density correlation functions $D(r)$ 
show exponential decay for various $V$, and 
(f) the charge structure factors $S_\textrm{c}(k)$ in 
the SDW phase show clearly quadratic behaviors at 
small $k$. (g) The singlet- and (h) triplet-pairing
correlation functions $\Phi_\textrm{S}(r)$ 
and $\Phi_\textrm{T}(r)$ both decay exponentially, 
and are both extremely short-ranged at $V=-2.44$ 
in the PS$_2$ phase. In practical calculations, 
for the calculations of $G(r)$, $F(r)$ and $D(r)$ 
in (a,c,e), the reference site is set at $L/2$; 
while for pairing correlations $\Phi_\textrm{S}$ 
and $\Phi_\textrm{T}$ in (g,h) the reference point 
is at $L/4+1$.
}
\label{Fig:SupportU4}
\end{figure}

\textbf{The $U=4$ case.} 
In the large $U=4$ case, there exists only SDW and PS$_2$ 
but no intermediate TS phase. The transition point to 
PS$_2$ can be accurately determined by the method described 
in \ref{SN:DMRG} above (c.f.,
\Fig{Fig:DetermineVPS2}), and it is estimated as 
$V_\textrm{s} \simeq -2.42$. In \Fig{Fig:SupportU4} we plot 
various correlation functions and structure factors 
throughout $V_\textrm{s} < V \le 0$. Except the quasi-long-range 
AFM correlation with scaling $(-1)^r/r$ in 
\Fig{Fig:SupportU4}(c), the other correlation functions
including the single-particle Green's function
[\Fig{Fig:SupportU4}(a)], the charge correlation
[\Fig{Fig:SupportU4}(e)], as well as the singlet and 
triplet pairing correlations [\Fig{Fig:SupportU4}(g,h)], 
all decay exponentially. In TLL, although there is no true 
Fermi point existing, the momentum-space distribution $n(k)$
nevertheless exhibits singularity at $\pm k_F$
\cite{Voit_RepProgPhys1995,Giamarchi1D,FradkinFieldTheory}.
In \Fig{Fig:SupportU4}(b) we see the $n(k)$ curves are 
smooth and there is no singularity at $k=\pm k_F=\pm\pi/4$, 
thus showing the absence of TLL states. Moreover, 
the charge structure factor $S_\textrm{c}(k)$ 
[see \Fig{Fig:SupportU4}(f)] around $k=0$ is quadratic, 
which indicates $K_\rho=0$. Regarding the spin
structure factor $S_\textrm{m}(k)$ in \Fig{Fig:SupportU4}(d), 
it diverges at $k=\pi$ and is consistent with the 
$(-1)^r/r$ scaling of $F(r)$ in \Fig{Fig:SupportU4}(c). 
To sum up, throughout $V_\textrm{s} < V \le 0$ the charge-related
excitations are fully gapped while the spin excitation is
gapless at $k=\pi$. These calculations confirm for $U=4$ 
there is no intermediate TS phase between the SDW and 
PS$_2$ phases.
\\

\subsection{More DMRG Results at Quarter Filling}
\label{SN:QF}

\begin{figure}[!tbp]
\includegraphics[width=1\linewidth]{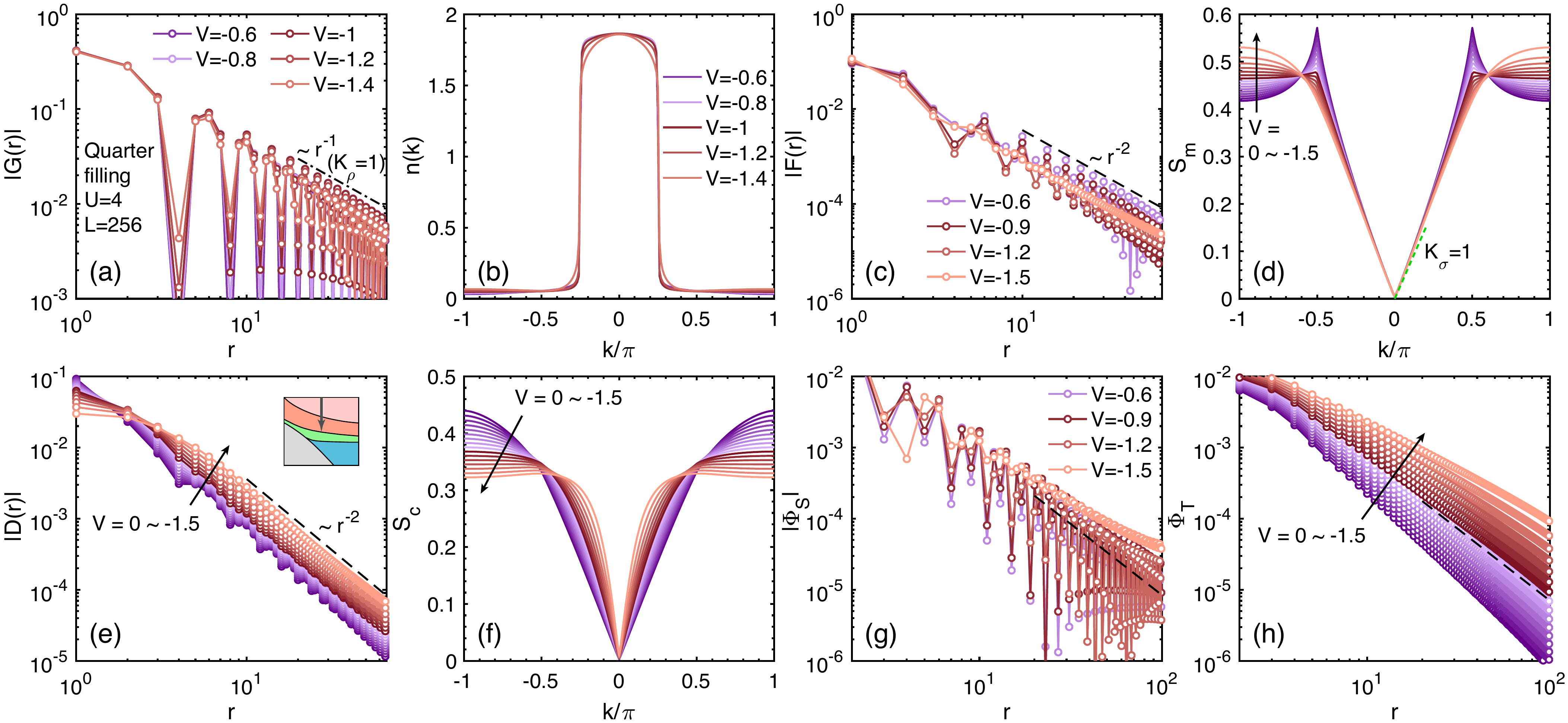}
\renewcommand{\figurename}{\textbf{Supplementary Figure}}
\caption{\textbf{Various correlation functions and structure 
factors of the EHM at quarter filling.} With a fixed $U=4$, 
the interactions $V$ are scanned through the uniform phases 
TLL and TS. (a) shows the single-particle Green's function
$G(r)$, where the dotted dashed line represents the 
$~r^{-1}$ scaling that corresponds to $K_\rho=1$. 
(b) The particle number distribution $n(k)$ in  
momentum space, and there exist singularities at 
$k = \pm k_F = \pm \pi/4$ where the derivatives of 
$n(k)$ vs $k$ diverge. (c) The spin correlation function, 
where the dashed line, as well as those in (e,g,h) 
denote the $~r^{-2}$ power-law behavior. (d) The spin 
structure factor $S_\textrm{m}(k)$, whose singularity at 
$k = \pm k_F = \pm \pi/2$ weakens as $|V|$ enhances. 
The dashed blue line denotes the linear relation 
reflecting $K_\sigma=1$. (e) Charge density correlation 
functions $D(r)$, (f) charge structure factors $S_\textrm{c}(k)$,
(g) singlet pairing correlation $\Phi_\textrm{S}(r)$, and
(f) the triplet pairing correlation $\Phi_\textrm{T}(r)$ 
are shown. Both paring correlations are enhanced as $|V|$ 
increases and the $\Phi_\textrm{T}(r)$ dominates in the charge-2e 
channel. In the calculations of $G(r)$, $F(r)$ and $D(r)$ 
in (a,c,e), the reference site is set at $L/2$, 
while for $\Phi_\textrm{S}$ and $\Phi_\textrm{T}$ 
in (g,h) the reference point is set at $L/4+1$. 
{In the plot we {have changed} the color code of the TLL 
phase from light red in the main text to the purple color
for the sake of visibility.} 
}
\label{Fig:SupportQF}
\end{figure}

Below we provide supportive data for determination of 
the ground-state phase diagram of EHM at quarter filling.\\ 

\textbf{Various correlations and structure factors at quarter filling.} 
In \Fig{Fig:SupportQF}, we consider the case of $U=4$ 
and show various correlation and structure factor results.
Similar to the half-filling case in \Fig{Fig:SupportU4},
here we again have $K_\sigma=1$ for the spin gapless cases.
At the crossover point $V_\textrm{c}\simeq -0.8$ we have $K_\rho=1$ 
and the $G(r)$ decays algebraically with a power of $-1$ [c.f.
Eq.~(\ref{EqS:GF})]. Moreover, for $V<V_\textrm{c}$ it still decays in 
power-law but the exponent becomes smaller than $-1$ as 
$K_\rho>1$. In \Fig{Fig:SupportQF}(b) the momentum-space 
particle number distribution $n(k)$ shows singularity at $\pm \pi/4$ 
(i.e., $\pm k_F$) for different interactions $V$ within the TLL 
and TS phases.

For the spin and charge correlations in \Fig{Fig:SupportQF}(c-f), 
we find there exists weak $2k_F$ singularity in the TLL 
phase, which becomes very weak (and even negligible) in the
TS phases [c.f., \Fig{Fig:SupportQF}(c,e)]. The 2$k_F$
singularity can be clearly seen in the spin structure factor
results shown in \Fig{Fig:SupportQF}(d), where the
structural peak at $\pm\pi/2$ rather prominent in the TLL 
phase becomes smeared out in the TS regime. This is due to 
the fact that the dominant correlations in both the spin 
and charge channels in the TS phase belong to the uniform 
$k=0$ mode.

Lastly, in \Fig{Fig:SupportQF}(g,h) we show the singlet and 
triplet pairing correlations, both of which exhibit algebraic behaviors 
in the TLL and TS phases. It can be noticed that $\Phi_\textrm{T}$ gets 
significantly enhanced as $|V|$ increases in the TS phase, 
where the power-law exponent exceeds $-2$ (meaning decaying 
more slowly), while the intensity of $\Phi_\textrm{S}$ barely changes as 
$|V|$ enhances.
\\

\begin{figure}[!tbp]
\includegraphics[width=0.9\linewidth]{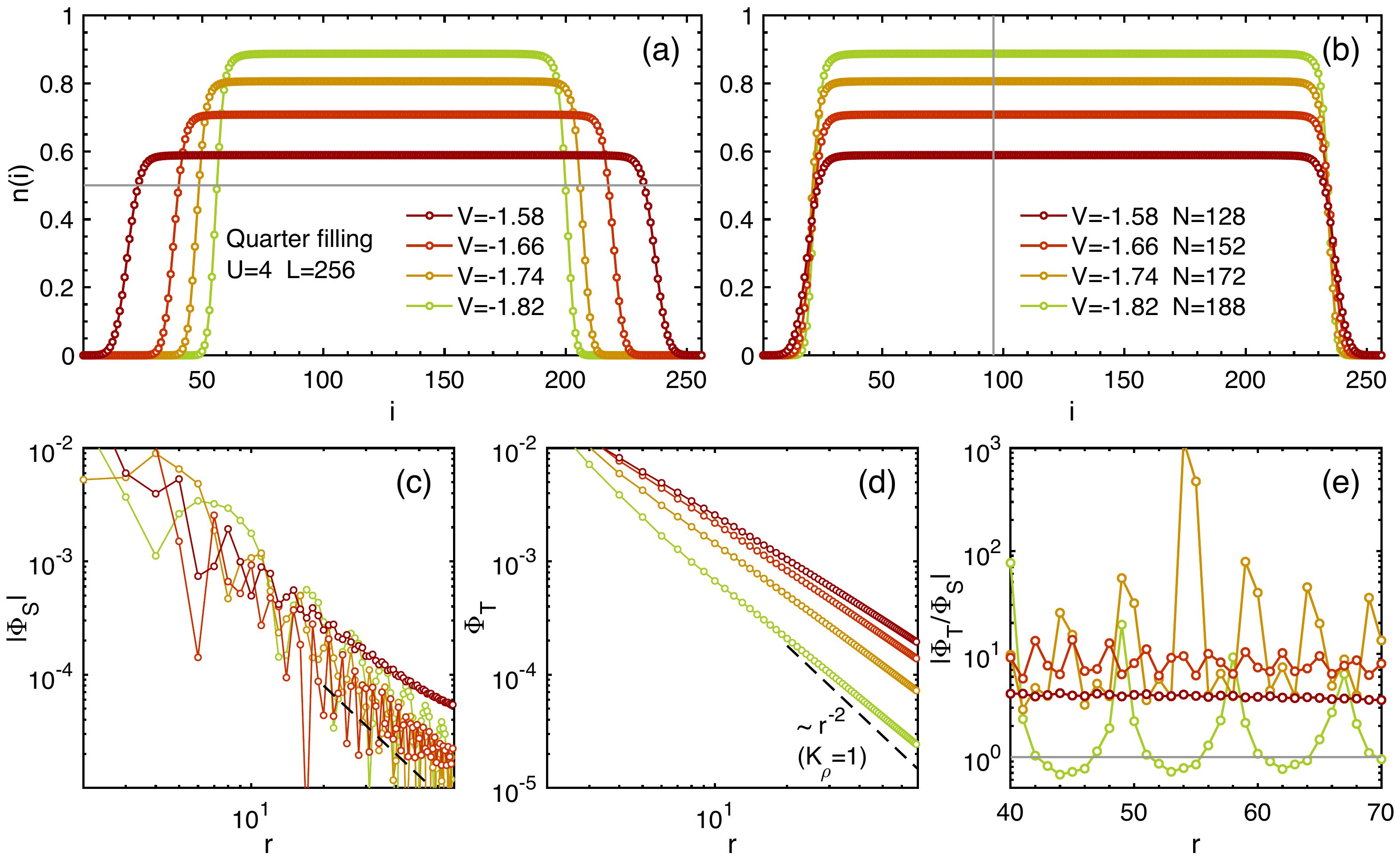}
\renewcommand{\figurename}{\textbf{Supplementary Figure}}
\caption{\textbf{Superconductivity pairing in the clustered plateau 
of the PS$_x$ phase.} The calculations are performed on the 
EHM with a fixed $U=4$ and various attractive interactions $V$ 
[see the legend in (b)]. {(a) Charge density distribution $n(i)$ in 
PS$_x$ at quarter filling, where the plateaus represent the 
clustered electrons. The horizontal gray line denotes the quarter 
filling. (b) $n(i)$ for the same interactions $U$ and $V$ in (a)
while with more total electron number $N\geq128$. The clusters
expand in their ranges but preserve the densities (as well as
electronic states and properties) in the plateau regimes. The 
vertical gray line indicates the reference site $i=3L/8$ in 
calculating the correlation functions.} (c) and (d) are the 
singlet ($\Phi_\textrm{S}$) and triplet ($\Phi_\textrm{T}$) pairing correlation 
functions, respectively, where the dashed line represent the 
$r^{-2}$ power-law scaling corresponding to $K_\rho=1$. 
(e) shows the ratio between the triplet and singlet pairing 
correlations $|\Phi_\textrm{T}/\Phi_\textrm{S}|$, which clearly exceeds 1 
(the gray line) at long distances.
}
\label{Fig:PSxSC}
\end{figure}

\textbf{Pairing correlations in the clustered region of PS$_x$.}
In the main text, we have mentioned the clustered electrons 
in PS$_x$ constitute a gapless TLL state, which even has
prominent TS pairing in a parameter regime not far from 
the uniform TS regime (c.f., the regime with $K_\rho>1$ in 
Fig.~5\textbf{a} of the main text). To be more specific, 
in \Fig{Fig:PSxSC}(c,d) we show respectively the 
singlet and triplet pairing correlation functions $\Phi_\textrm{S}$ 
and $\Phi_\textrm{T}$ measured in the clustered part [i.e., plateaus 
in \Fig{Fig:PSxSC}(a,b)]. 

Interestingly, we can further add some electrons to the 
system and make the whole system away from quarter filling,
which, however, does not alter the electronic states in 
the clustered plateaus [see \Fig{Fig:PSxSC}(b)]. When 
computing the pairing correlations in these wider plateaus, 
the boundary effects are further reduced, and the results 
in \Fig{Fig:PSxSC}(c,d,e) are in practice computed in this
manner. The results confirm that the properties of the 
clusters depend only on $U$,$V$ but not on the total filling
factor.

In \Fig{Fig:PSxSC}(c,d), we find both pairing 
correlations decay slower than $r^{-2}$ in the electron
clustered regime, and the triplet pairing $\Phi_\textrm{T}$ again
dominates. In \Fig{Fig:PSxSC}(e) we show the ratio
$|\Phi_\textrm{T}/\Phi_\textrm{S}|$ directly, and find at long distances 
$\Phi_\textrm{T}$ is significantly stronger than $\Phi_\textrm{S}$, confirming
the TS nature of the gapless clustered electrons.


\end{document}